\documentclass{ws-ijgmmp}
\usepackage[utf8]{inputenc}
\usepackage[T1]{fontenc}
\usepackage[colorlinks,hyperindex,plainpages=false]{hyperref}
\usepackage{cite}
\usepackage{amsmath}
\usepackage{amssymb}
\usepackage{amsfonts}
\usepackage{accents}
\allowdisplaybreaks

\newcommand{\dd}{\mathrm{d}}

\begin{document}

\markboth{Manuel Hohmann}
{Spatially homogeneous teleparallel spacetimes with four-dimensional groups of motions}

%
\catchline{}{}{}{}{}
%

\title{Spatially homogeneous teleparallel spacetimes with four-dimensional groups of motions}

\author{Manuel Hohmann}

\address{Laboratory of Theoretical Physics, Institute of Physics, University of Tartu, W. Ostwaldi 1, 50411 Tartu, Estonia\\
\email{manuel.hohmann@ut.ee}}

\maketitle

\begin{history}
\received{(Day Month Year)}
\revised{(Day Month Year)}
\end{history}

\begin{abstract}
We study metric teleparallel geometries, which can either be defined through a Lorentzian metric and flat, metric-compatible affine connection, or a tetrad and a flat Lorentz spin connection, which are invariant under the transitive action of a four-dimensional Lie group on their spatial equal-time hypersurfaces. There are three such group actions, and their corresponding spatial hypersurfaces belong to the Bianchi types II, III and IX, respectively. For each of these three symmetry groups, we determine the most general teleparallel geometry, and find that it is parametrized by six functions of time, one of which can be eliminated by the choice of the time coordinate. We further show that these geometries are unique up to global Lorentz transformations, coordinate transformations and changes of the choice of the parameter functions.
\end{abstract}



\section{Introduction}\label{sec:intro}
Current observations in cosmology, such as different measurements of the present-time value \(H_0\) of the Hubble parameter~\cite{Planck:2018vyg} by either early-time or late-time methods, put tensions on the so-called $\Lambda$CDM model, which models the dynamics of the universe through general relativity, a cosmological constant $\Lambda$ and cold dark matter (CDM), and is nowadays considered as the standard model of cosmology. Various potential explanations for these observations have been offered, such as introducing new and unknown types of matter, or modifications of the underlying theory of gravity, and their implications on cosmological observations have been investigated~\cite{DiValentino:2021izs,CANTATA:2021ktz,Heisenberg:2018vsk}. These modifications include a class of theories in which gravity is not attributed to the curvature of the Levi-Civita connection, but to the torsion of a flat, metric-compatible connection, known as teleparallel gravity~\cite{Aldrovandi:2013wha,Golovnev:2018red,Hohmann:2022mlc}, their homogeneous and isotropic backgrounds have been derived~\cite{Hohmann:2020zre,Coley:2022qug} and the resulting cosmology has been extensively studied~\cite{Bahamonde:2021gfp,Cai:2015emx}.

An important tool to study cosmology and relate gravity theories to cosmological observables is (predominantly linear) cosmological perturbation theory~\cite{Kodama:1984ziu,Mukhanov:1990me,Malik:2008im}. Its fundamental assumption is to split the full dynamics of matter and the gravitational field into a homogeneous and isotropic background, and a non-homogeneous, non-isotropic perturbation. While this method is applicable without restrictions to a large number of modified gravity theories, it turns out that in teleparallel gravity theories a so-called strong coupling problem arises. One symptom of strong coupling is the fact that certain modes in the linear perturbation theory become non-dynamical for highly symmetric background solutions, such as the homogeneous and isotropic background conventionally assumed in cosmology~\cite{Golovnev:2018wbh,Golovnev:2020zpv,BeltranJimenez:2020fvy,Blagojevic:2020dyq,Guzman:2019oth,BeltranJimenez:2019nns,Bahamonde:2022ohm}. Several possible solutions to this problem have been proposed, such as a relaxed notion of symmetry for teleparallel spacetimes, which applies the homogeneity and isotropy only to the metric, but allows the flat connection to possess less symmetry~\cite{Golovnev:2020nln,Li:2023fto}.

In this article we retain the notion of symmetry, demanding that both the metric and the connection are invariant under the full symmetry group, but we consider smaller symmetry groups than conventionally assumed in cosmology. In particular, we do not impose the symmetry under the full isotropy group \(\mathrm{SO}(3)\) at every spacetime point, but keep the demand of spatial homogeneity, i.e., the existence of a transitive action of a group \(G\) on the spatial hypersurfaces which leaves the teleparallel geometry invariant. The dynamics of such homogeneous, anisotropic teleparallel cosmologies have been studied, e.g., in~\cite{Rodrigues:2012qua,Rodrigues:2014xam,Paliathanasis:2022vux,Amir:2015wja,Sharif:2011bi,Fayaz:2014swa,Fayaz:2015yka,Skugoreva:2017vde,Skugoreva:2019bwt,Tretyakov:2021cgb,Rodrigues:2013iua,Aslam:2013coa,Paliathanasis:2016vsw,Paliathanasis:2017htk,Coley:2023ibm}. The possible transitive group actions on three-dimensional Riemannian manifolds have been classified by Bianchi~\cite{Bianchi:1898,Bianchi:2001}, who has shown that among the nine classes there are three classes whose invariant metric allows for a four-dimensional symmetry group. These so-called Bianchi types II, III and IX are the starting point of the construction we present here. In this article we construct the most general teleparallel geometries which are invariant under the action of the respective four-dimensional groups found by Bianchi.

For our calculation, we make use of a formulation of teleparallel spacetime symmetries which demands the invariance of both the metric and the teleparallel affine connection under the action of the group \(G\), and relates them to the invariance of the fundamental field variables, a tetrad and a spin connection, up to a local Lorentz transformation~\cite{Hohmann:2019nat,McNutt:2023nxm}. An important ingredient in this formalism is a Lie algebra homomorphism \(\boldsymbol{\lambda}: \mathfrak{g} \to \mathfrak{so}(1,3)\) from the symmetry algebra \(\mathfrak{g}\) to the Lorentz algebra \(\mathfrak{so}(1,3)\). In order to find a symmetric tetrad and spin connection, one needs to specify a homomorphism, and in order to find all teleparallel geometries possessing the demanded symmetry, one needs to construct all homomorphisms. An important part of the work we present here is therefore the construction of all algebra homomorphisms for the four-dimensional symmetry algebras underlying the aforementioned Bianchi spacetimes. For this purpose, we provide a simple, constructive, geometric approach. Once we have found all homomorphisms, we use them to derive all teleparallel geometries obeying the corresponding symmetries.

This article is structured as follows. In section~\ref{sec:symtele} we provide a brief review of the notion of spacetime symmetries in teleparallel gravity. We then turn our focus to the Lorentz algebra in section~\ref{sec:lorentz}, where we provide a simple geometric representation of the algebra and study some of its properties, in particular the orbits of the adjoint group action. We then use this formalism in order to construct all Lie algebra homomorphisms from the four-dimensional symmetry algebras to the Lorentz algebra in section~\ref{sec:algebra}. In section~\ref{sec:cosmo} we then use these homomorphisms in order to construct the resulting teleparallel geometries. We end with a conclusion in section~\ref{sec:conclusion}.

\section{Teleparallel geometry with spacetimes symmetries}\label{sec:symtele}
We begin our discussion with a brief review of the notion of metric teleparallel geometries and their spacetime symmetries. The dynamical field variables, which determine the teleparallel geometry, are the tetrad \(\theta^a = \theta^a{}_{\mu}\dd x^{\mu}\) and the spin connection \(\omega^a{}_b = \omega^a{}_{b\mu}\dd x^{\mu}\). The former is assumed to be invertible, with inverse \(e_a{}^{\mu}\) satisfying
\begin{equation}
\theta^a{}_{\mu}e_a{}^{\nu} = \delta_{\mu}^{\nu}\,, \quad
\theta^a{}_{\mu}e_b{}^{\mu} = \delta^a_b\,,
\end{equation}
while the latter is antisymmetric,
\begin{equation}
\eta_{ac}\omega^c{}_{b\mu} + \eta_{bc}\omega^c{}_{a\mu} = 0\,,
\end{equation}
and flat,
\begin{equation}
\partial_{\mu}\omega^a{}_{b\nu} - \partial_{\nu}\omega^a{}_{b\mu} + \omega^a{}_{c\mu}\omega^c{}_{b\nu} - \omega^a{}_{c\nu}\omega^c{}_{b\mu} = 0\,.
\end{equation}
Together they define the metric
\begin{equation}\label{eq:metric}
g_{\mu\nu} = \eta_{ab}\theta^a{}_{\mu}\theta^b{}_{\nu}
\end{equation}
and the teleparallel affine connection
\begin{equation}\label{eq:affcon}
\Gamma^{\mu}{}_{\nu\rho} = e_a{}^{\mu}(\partial_{\rho}\theta^a{}_{\nu} + \omega^a{}_{b\rho}\theta^b{}_{\nu})\,.
\end{equation}
We say that a set of vector fields spanned by \(X_A = X_A^{\mu}\partial_{\mu}\) which is closed under the Lie bracket, and hence forms a Lie algebra \(\mathfrak{g}\) with structure constants \(C_{AB}{}^C\) defined by
\begin{equation}
[X_A, X_B] = C_{AB}{}^CX_C
\end{equation}
acting on the spacetime manifold \(M\), generates a \emph{symmetry} of a teleparallel geometry if it leaves the metric and the teleparallel connection invariant, i.e., if it satisfies
\begin{equation}\label{eq:metsymcond}
0 = (\mathcal{L}_{X_A}g)_{\mu\nu} = X_A^{\rho}\partial_{\rho}g_{\mu\nu} + \partial_{\mu}X_A^{\rho}g_{\rho\nu} + \partial_{\nu}X_A^{\rho}g_{\mu\rho}
\end{equation}
and
\begin{equation}\label{eq:affsymcond}
0 = (\mathcal{L}_{X_A}\Gamma)^{\mu}{}_{\nu\rho} = X_A^{\sigma}\partial_{\sigma}\Gamma^{\mu}{}_{\nu\rho} - \partial_{\sigma}X_A^{\mu}\Gamma^{\sigma}{}_{\nu\rho} + \partial_{\nu}X_A^{\sigma}\Gamma^{\mu}{}_{\sigma\rho} + \partial_{\rho}X_A^{\sigma}\Gamma^{\mu}{}_{\nu\sigma} + \partial_{\nu}\partial_{\rho}X_A^{\mu}\,.
\end{equation}
As shown in~\cite{Hohmann:2019nat}, this is the case if and only if there exists a map \(\boldsymbol{\lambda}: \mathfrak{g} \times M \to \mathfrak{so}(1,3)\) which is a local Lie algebra homomorphism, i.e., which satisfies
\begin{equation}
[\boldsymbol{\lambda}(X_A, x), \boldsymbol{\lambda}(X_B, x)] = \boldsymbol{\lambda}([X_A, X_B], x) = C_{AB}{}^C\boldsymbol{\lambda}(X_C, x)\,,
\end{equation}
such that
\begin{equation}\label{eq:infisymcondgen}
(\mathcal{L}_{X_A}\theta)^a{}_{\mu} = -\boldsymbol{\lambda}_A^a{}_b\theta^b{}_{\mu}\,, \quad
(\mathcal{L}_{X_A}\omega)^a{}_{b\mu} = \partial_{\mu}\boldsymbol{\lambda}_A^a{}_b + \omega^a{}_{c\mu}\boldsymbol{\lambda}_A^c{}_b - \omega^c{}_{b\mu}\boldsymbol{\lambda}_A^a{}_c\,,
\end{equation}
where we used the abbreviation \(\boldsymbol{\lambda}_A^a{}_b(x) = \boldsymbol{\lambda}^a{}_b(X_A, x)\). A particular property of the teleparallel geometry is the fact that the metric~\eqref{eq:metric} and the affine connection~\eqref{eq:affcon}, which appear in the teleparallel action and field equations, are invariant under local Lorentz transformations of the form
\begin{equation}
\theta^a{}_{\mu} \mapsto \Lambda^a{}_b\theta^b{}_{\mu}\,, \quad
\omega^a{}_{b\mu} \mapsto \Lambda^a{}_c\Lambda_b{}^d\omega^c{}_{d\mu} + \Lambda^a{}_c\partial_{\mu}\Lambda_b{}^c\,.
\end{equation}
Choosing a suitable Lorentz transformation, it is always possible to achieve the so-called Weitzenböck gauge, which is defined by the vanishing spin connection, \(\omega^a{}_{b\mu} \equiv 0\), at least locally~\cite{Hohmann:2021dhr}. In this case the symmetry conditions simplify and become
\begin{equation}\label{eq:infisymcondwb}
(\mathcal{L}_{X_A}\theta)^a{}_{\mu} = -\boldsymbol{\lambda}_A^a{}_b\theta^b{}_{\mu}\,, \quad
0 \equiv (\mathcal{L}_{X_A}\omega)^a{}_{b\mu} = \partial_{\mu}\boldsymbol{\lambda}_A^a{}_b\,.
\end{equation}
Note in particular that the second condition implies that \(\boldsymbol{\lambda}\) is independent of the spacetime point, and thus becomes an ordinary Lie algebra homomorphism \(\boldsymbol{\lambda}: \mathfrak{g} \to \mathfrak{so}(1,3)\). A crucial step in finding the most general teleparallel geometry which is symmetric under the action of a given Lie algebra \(\mathfrak{g}\) is to find all such Lie algebra homomorphisms. In the following section, we will present a method which turns out to be particularly useful for this task.

\section{Vector product representation of the Lorentz algebra}\label{sec:lorentz}
In order to simplify the construction and classification of all Lie algebra homomorphisms \(\boldsymbol{\lambda}: \mathfrak{g} \to \mathfrak{so}(1,3)\) to the extent possible, consider the basis of the Lorentz algebra \(\mathfrak{so}(1,3)\) given by the explicit matrix representation
\begin{subequations}\label{eq:loralgmat}
\begin{align}
J_1 &= \begin{pmatrix}
0 & 0 & 0 & 0\\
0 & 0 & 0 & 0\\
0 & 0 & 0 & -1\\
0 & 0 & 1 & 0
\end{pmatrix}\,, &
J_2 &= \begin{pmatrix}
0 & 0 & 0 & 0\\
0 & 0 & 0 & 1\\
0 & 0 & 0 & 0\\
0 & -1 & 0 & 0
\end{pmatrix}\,, &
J_3 &= \begin{pmatrix}
0 & 0 & 0 & 0\\
0 & 0 & -1 & 0\\
0 & 1 & 0 & 0\\
0 & 0 & 0 & 0
\end{pmatrix}\,,\\
K_1 &= \begin{pmatrix}
0 & 1 & 0 & 0\\
1 & 0 & 0 & 0\\
0 & 0 & 0 & 0\\
0 & 0 & 0 & 0
\end{pmatrix}\,, &
K_2 &= \begin{pmatrix}
0 & 0 & 1 & 0\\
0 & 0 & 0 & 0\\
1 & 0 & 0 & 0\\
0 & 0 & 0 & 0
\end{pmatrix}\,, &
K_3 &= \begin{pmatrix}
0 & 0 & 0 & 1\\
0 & 0 & 0 & 0\\
0 & 0 & 0 & 0\\
1 & 0 & 0 & 0
\end{pmatrix}\,.
\end{align}
\end{subequations}
These basis elements satisfy the well-known commutation relations
\begin{equation}\label{eq:lorcomm}
[J_i, J_j] = \epsilon_{ijk}J_k\,, \quad
[K_i, K_j] = -\epsilon_{ijk}J_k\,, \quad
[J_i, K_j] = \epsilon_{ijk}K_k\,.
\end{equation}
We then introduce the notation
\begin{equation}
\vec{J} = (J_1, J_2, J_3)\,, \quad
\vec{K} = (K_1, K_2, K_3)\,,
\end{equation}
so that we can write any element of \(\mathfrak{so}(1,3)\) in the form
\begin{equation}
\vec{j} \cdot \vec{J} + \vec{k} \cdot \vec{K}
\end{equation}
for \(\vec{j}, \vec{k} \in \mathbb{R}^3\). The commutation relations~\eqref{eq:lorcomm} are then equivalently written as
\begin{equation}
[\vec{j} \cdot \vec{J}, \vec{j}' \cdot \vec{J}] = (\vec{j} \times \vec{j}') \cdot \vec{J}\,, \quad
[\vec{k} \cdot \vec{K}, \vec{k}' \cdot \vec{K}] = -(\vec{k} \times \vec{k}') \cdot \vec{J}\,, \quad
[\vec{j} \cdot \vec{J}, \vec{k} \cdot \vec{K}] = (\vec{j} \times \vec{k}) \cdot \vec{K}\,,
\end{equation}
using the vector product. For two arbitrary elements of the Lorentz algebra we thus have
\begin{equation}
[\vec{j} \cdot \vec{J} + \vec{k} \cdot \vec{K}, \vec{j}' \cdot \vec{J} + \vec{k}' \cdot \vec{K}] = (\vec{j} \times \vec{j}' - \vec{k} \times \vec{k}') \cdot \vec{J} + (\vec{j} \times \vec{k}' + \vec{k} \times \vec{j}') \cdot \vec{K}\,,
\end{equation}
showing that they resemble the vector product of complex vectors \(\vec{j} + i\vec{k} \in \mathbb{C}^3\). In the following, however, it will turn out to be simpler to treat them as pairs of real vectors. In order to understand the coordinate freedom given by conjugacy classes of Lie algebra homomorphisms, it is further helpful to calculate the adjoint group representation. For this purpose, let \(\vec{n}\) be a unit vector. By direct calculation one finds
\begin{subequations}
\begin{align}
e^{t\vec{n} \cdot \vec{J}}(\vec{j} \cdot \vec{J})e^{-t\vec{n} \cdot \vec{J}} &= [(\vec{n} \cdot \vec{j})\vec{n} + \vec{n} \times \vec{j}\sin t - \vec{n} \times (\vec{n} \times \vec{j})\cos t] \cdot \vec{J}\,,\\
e^{t\vec{n} \cdot \vec{J}}(\vec{k} \cdot \vec{K})e^{-t\vec{n} \cdot \vec{J}} &= [(\vec{n} \cdot \vec{k})\vec{n} + \vec{n} \times \vec{k}\sin t - \vec{n} \times (\vec{n} \times \vec{k})\cos t] \cdot \vec{K}\,,\\
e^{t\vec{n} \cdot \vec{K}}(\vec{j} \cdot \vec{J})e^{-t\vec{n} \cdot \vec{K}} &= [(\vec{n} \cdot \vec{j})\vec{n} - \vec{n} \times (\vec{n} \times \vec{j})\cosh t] \cdot \vec{J} + \vec{n} \times \vec{j} \cdot \vec{K}\sinh t\,,\\
e^{t\vec{n} \cdot \vec{K}}(\vec{k} \cdot \vec{K})e^{-t\vec{n} \cdot \vec{K}} &= [(\vec{n} \cdot \vec{k})\vec{n} - \vec{n} \times (\vec{n} \times \vec{k})\cosh t] \cdot \vec{K} - \vec{n} \times \vec{k} \cdot \vec{J}\sinh t\,.
\end{align}
\end{subequations}
Note that this transformation leaves the Killing form
\begin{equation}
\kappa(\vec{j} \cdot \vec{J} + \vec{k} \cdot \vec{K}, \vec{j}' \cdot \vec{J} + \vec{k}' \cdot \vec{K}) = 4(\vec{k} \cdot \vec{k'} - \vec{j} \cdot \vec{j}')
\end{equation}
invariant by construction. In addition, one finds that also the bilinear form
\begin{equation}
\iota(\vec{j} \cdot \vec{J} + \vec{k} \cdot \vec{K}, \vec{j}' \cdot \vec{J} + \vec{k}' \cdot \vec{K}) = \vec{j} \cdot \vec{k'} + \vec{k} \cdot \vec{j}'
\end{equation}
is invariant. This allows us to classify the orbits of this action, which will turn out to be useful later. First note that rotations preserve the lengths \(\|\vec{j}\|, \|\vec{k}\| \in [0, \infty)\) as well as the angle \(\sphericalangle(\vec{j}, \vec{k}) \in [0, \pi]\), where the latter can conveniently be expressed by its cosine. Any orbit under rotations is uniquely expressed by these three quantities, or by the length(s) only if at least one of the two vectors is the zero vector. If one considers boosts, only the combinations
\begin{equation}\label{eq:orbinv}
S = \vec{j} \cdot \vec{k} = \|\vec{j}\|\|\vec{k}\|\cos\sphericalangle(\vec{j}, \vec{k})\,, \quad
D = \|\vec{j}\|^2 - \|\vec{k}\|^2
\end{equation}
are preserved. One then finds the following orbits:
\begin{enumerate}
\item
The orbit of the zero element \(\vec{j} = \vec{k} = 0\) has \(S = D = 0\) and contains only this element.
\item
There exists a second orbit with \(S = D = 0\) given by \(\|\vec{j}\| = \|\vec{k}\| > 0\) and \(\vec{j} \perp \vec{k}\). Boosts act by rescaling both vectors with a common factor.
\item
For any positive number \(c > 0\) there exists an orbit generated by the element \(\|\vec{j}\| = c\) and \(\vec{k} = 0\). The elements of this orbit have \(D = c^2 > 0\) and \(S = 0\) and consist of those pairs of vectors for which \(\vec{j} \perp \vec{k}\) and \(\|\vec{j}\|^2 = \|\vec{k}\|^2 + c^2\).
\item
Similarly to the previous case, there exists another orbit for any positive number \(c > 0\), which is generated by the element \(\|\vec{k}\| = c\) and \(\vec{j} = 0\). In this case one has \(D = -c^2 < 0\) and \(S = 0\). The elements of this orbit are those pairs of vectors for which \(\vec{j} \perp \vec{k}\) and \(\|\vec{k}\|^2 = \|\vec{j}\|^2 + c^2\).
\item
For any pair \(a, b > 0\) of positive numbers there exists the orbit of the element \(\|\vec{j}\| = a\) and \(\|\vec{k}\| = b\) with \(\sphericalangle(\vec{j}, \vec{k}) = 0\). This orbit has \(D = a^2 - b^2\) and \(S = ab\). It can be parametrized by \(C = \cos\sphericalangle(\vec{j}, \vec{k}) \in (0, 1]\) and one has
\begin{equation}
\|\vec{j}\|^2 = \frac{\sqrt{C^2D^2 + 4S^2} + CD}{2C}\,, \quad
\|\vec{k}\|^2 = \frac{\sqrt{C^2D^2 + 4S^2} - CD}{2C}\,.
\end{equation}
In particular, one has
\begin{equation}\label{eq:hyporbpar}
a^2 = \frac{\sqrt{D^2 + 4S^2} + D}{2}\,, \quad
b^2 = \frac{\sqrt{D^2 + 4S^2} - D}{2}\,,
\end{equation}
and all orbits with \(S > 0\) and \(D \in \mathbb{R}\) are of this type.
\item
Finally, all orbits with \(S < 0\) and \(D \in \mathbb{R}\) are characterized by the same numbers \(a, b > 0\) given by~\eqref{eq:hyporbpar}, and are generated by the element \(\|\vec{j}\| = a\) and \(\|\vec{k}\| = b\) with \(\sphericalangle(\vec{j}, \vec{k}) = \pi\). The orbit is now parametrized by
\begin{equation}
\|\vec{j}\|^2 = -\frac{\sqrt{C^2D^2 + 4S^2} - CD}{2C}\,, \quad
\|\vec{k}\|^2 = -\frac{\sqrt{C^2D^2 + 4S^2} + CD}{2C}\,.
\end{equation}
where \(C = \cos\sphericalangle(\vec{j}, \vec{k}) \in [-1, 0)\)\,.
\end{enumerate}
We can make use of the adjoint group action to transform any algebra homomorphism into a convenient form. Given a basis \((X_A)\) of the Lie algebra \(\mathfrak{g}\), we will use the notation
\begin{equation}
\boldsymbol{\lambda}(X_A) = \vec{j}_A \cdot \vec{J} + \vec{k}_A \cdot \vec{K}
\end{equation}
for the Lie algebra homomorphism \(\boldsymbol{\lambda}\). The condition that \(\boldsymbol{\lambda}\) is a Lie algebra homomorphism then reads
\begin{equation}
\vec{j}_A \times \vec{j}_B - \vec{k}_A \times \vec{k}_B = C_{AB}{}^C\vec{j}_C\,, \quad
\vec{j}_A \times \vec{k}_B + \vec{k}_A \times \vec{j}_B = C_{AB}{}^C\vec{k}_C\,.
\end{equation}
We are thus left with the task of finding the most general set of vectors \(\vec{j}_A, \vec{k}_A\) which satisfy these relations for a given set of structure constants \(C_{AB}{}^C\) defined  by the Lie algebra structure of \(\mathfrak{g}\). We will solve this task in the following section.

\section{Algebra homomorphisms}\label{sec:algebra}
We now make use of the vector representation introduced in the previous section in order to determine all (conjugacy classes of) Lie algebra homomorphisms from the symmetry algebra of the teleparallel spacetime to the Lorentz algebra. We do so for each Bianchi type under consideration. We discuss Bianchi type II in section~\ref{ssec:algebraII}, Bianchi type III in section~\ref{ssec:algebraIII} and Bianchi type IX in section~\ref{ssec:algebraIX}.

\subsection{Bianchi type II}\label{ssec:algebraII}
For the Bianchi type II, the three translation generators satisfy the relations
\begin{equation}\label{eq:algII123}
[X_1, X_2] = [X_1, X_3] = 0\,, \quad
[X_2, X_3] = X_1\,.
\end{equation}
Further, the general homogeneous metric admits another symmetry under a vector field \(X_4\) which satisfies
\begin{equation}\label{eq:algII4}
[X_1, X_4] = 0\,, \quad
[X_2, X_4] = -X_3\,, \quad
[X_3, X_4] = X_2\,.
\end{equation}
From these relations follows that the Lie algebra homomorphisms we aim to find must satisfy
\begin{subequations}\label{eq:vekII}
\begin{align}
\vec{j}_1 \times \vec{j}_2 - \vec{k}_1 \times \vec{k}_2 &= 0\,, &
\vec{j}_1 \times \vec{k}_2 + \vec{k}_1 \times \vec{j}_2 &= 0\,,\label{eq:vekIIa}\\
\vec{j}_1 \times \vec{j}_3 - \vec{k}_1 \times \vec{k}_3 &= 0\,, &
\vec{j}_1 \times \vec{k}_3 + \vec{k}_1 \times \vec{j}_3 &= 0\,,\label{eq:vekIIb}\\
\vec{j}_1 \times \vec{j}_4 - \vec{k}_1 \times \vec{k}_4 &= 0\,, &
\vec{j}_1 \times \vec{k}_4 + \vec{k}_1 \times \vec{j}_4 &= 0\,,\label{eq:vekIIc}\\
\vec{j}_2 \times \vec{j}_3 - \vec{k}_2 \times \vec{k}_3 &= \vec{j}_1\,, &
\vec{j}_2 \times \vec{k}_3 + \vec{k}_2 \times \vec{j}_3 &= \vec{k}_1\,,\label{eq:vekIId}\\
\vec{j}_2 \times \vec{j}_4 - \vec{k}_2 \times \vec{k}_4 &= -\vec{j}_3\,, &
\vec{j}_2 \times \vec{k}_4 + \vec{k}_2 \times \vec{j}_4 &= -\vec{k}_3\,,\label{eq:vekIIe}\\
\vec{j}_3 \times \vec{j}_4 - \vec{k}_3 \times \vec{k}_4 &= \vec{j}_2\,, &
\vec{j}_3 \times \vec{k}_4 + \vec{k}_3 \times \vec{j}_4 &= \vec{k}_2\,.\label{eq:vekIIf}
\end{align}
\end{subequations}
We see that we can explicitly solve the last pair~\eqref{eq:vekIIf} of equations for \(\vec{j}_2\) and \(\vec{k}_2\), and then substitute into the remaining equations. In the next step, we solve the fourth pair~\eqref{eq:vekIId} of equations for \(\vec{j}_1\) and \(\vec{k}_1\), and substitute once again. We are then left with system of equations for the remaining vectors \(\vec{j}_3, \vec{j}_4, \vec{k}_3, \vec{k}_4\). To proceed further, we make use of the fifth pair~\eqref{eq:vekIIe} of equations, for which we find, after substituting \(\vec{j}_2\) and \(\vec{k}_2\),
\begin{subequations}
\begin{align}
0 = (\vec{j}_2 \times \vec{j}_4 - \vec{k}_2 \times \vec{k}_4 + \vec{j}_3) \cdot \vec{k}_4 + (\vec{j}_2 \times \vec{k}_4 + \vec{k}_2 \times \vec{j}_4 + \vec{k}_3) \cdot \vec{j}_4 &= \vec{j}_3 \cdot \vec{k}_4 + \vec{j}_4 \cdot \vec{k}_3\,,\\
0 = (\vec{j}_2 \times \vec{j}_4 - \vec{k}_2 \times \vec{k}_4 + \vec{j}_3) \cdot \vec{j}_4 - (\vec{j}_2 \times \vec{k}_4 + \vec{k}_2 \times \vec{j}_4 + \vec{k}_3) \cdot \vec{k}_4 &= \vec{j}_3 \cdot \vec{j}_4 - \vec{k}_3 \cdot \vec{k}_4\,.
\end{align}
\end{subequations}
Hence, we obtain the conditions
\begin{equation}\label{eq:condII34}
\vec{j}_3 \cdot \vec{k}_4 + \vec{j}_4 \cdot \vec{k}_3 = \vec{j}_3 \cdot \vec{j}_4 - \vec{k}_3 \cdot \vec{k}_4 = 0\,.
\end{equation}
Solving both conditions for their second terms and substituting back into the equations~\eqref{eq:vekIIe}, they greatly simplify and become
\begin{subequations}
\begin{align}
(\vec{j}_4 \cdot \vec{j}_4 - \vec{k}_4 \cdot \vec{k}_4 - 1)\vec{j}_3 - 2(\vec{j}_4 \cdot \vec{k}_4)\vec{k}_3 &= 0\,,\\
(\vec{j}_4 \cdot \vec{j}_4 - \vec{k}_4 \cdot \vec{k}_4 - 1)\vec{k}_3 + 2(\vec{j}_4 \cdot \vec{k}_4)\vec{j}_3 &= 0\,.
\end{align}
\end{subequations}
Taking the scalar product of these equations with \(\vec{j}_3\) and \(\vec{k}_3\), or vice versa, and calculating their sum, or difference, respectively, we find the equations
\begin{equation}
2(\vec{j}_3 \cdot \vec{j}_3 + \vec{k}_3 \cdot \vec{k}_3)(\vec{j}_4 \cdot \vec{k}_4) = (\vec{j}_3 \cdot \vec{j}_3 + \vec{k}_3 \cdot \vec{k}_3)(\vec{j}_4 \cdot \vec{j}_4 - \vec{k}_4 \cdot \vec{k}_4 - 1) = 0\,.
\end{equation}
There are two possibilities to solve these equations:
\begin{enumerate}
\item
The common factor vanishes if and only if \(\vec{j}_3 = \vec{k}_3 = 0\). In this case it follows from the full equations~\eqref{eq:vekII} that also \(\vec{j}_1 = \vec{j}_2 = \vec{k}_1 = \vec{k}_2 = 0\). All equations are satisfied under these conditions, and there are no restrictions on \(\vec{j}_4\) and \(\vec{k}_4\).

\item
Alternatively, the equations are solved by
\begin{equation}\label{eq:condII4}
\vec{j}_4 \cdot \vec{k}_4 = \vec{j}_4 \cdot \vec{j}_4 - \vec{k}_4 \cdot \vec{k}_4 - 1 = 0\,.
\end{equation}
In this case one finds a non-trivial homomorphism.
\end{enumerate}
To further proceed with the second class of solutions, we now consider the pair~\eqref{eq:vekIIb}. Substituting all previously found conditions, these become
\begin{subequations}
\begin{align}
(\vec{j}_3 \cdot \vec{j}_3 - \vec{k}_3 \cdot \vec{k}_3)\vec{j}_3 - 2(\vec{j}_3 \cdot \vec{k}_3)\vec{k}_3 &= 0\,,\\
(\vec{j}_3 \cdot \vec{j}_3 - \vec{k}_3 \cdot \vec{k}_3)\vec{k}_3 + 2(\vec{j}_3 \cdot \vec{k}_3)\vec{j}_3 &= 0\,.
\end{align}
\end{subequations}
Once again, we can take the scalar product of these equations with \(\vec{j}_3\) and \(\vec{k}_3\), or vice versa, and calculate their sum, or difference, respectively. This leads to the equations
\begin{equation}
(\vec{j}_3 \cdot \vec{j}_3)^2 - (\vec{k}_3 \cdot \vec{k}_3)^2 = 2(\vec{j}_3 \cdot \vec{j}_3 + \vec{k}_3 \cdot \vec{k}_3)(\vec{j}_3 \cdot \vec{k}_3) = 0\,.
\end{equation}
Besides the trivial solution \(\vec{j}_3 = \vec{k}_3 = 0\), they are solved by the conditions
\begin{equation}\label{eq:condII3}
\vec{j}_3 \cdot \vec{j}_3 - \vec{k}_3 \cdot \vec{k}_3 = \vec{j}_3 \cdot \vec{k}_3 = 0\,.
\end{equation}
Finally, substituting all conditions found so far into the first pair~\eqref{eq:vekIIa} of equations, we find that these are solved identically. Hence, no further conditions arise, and so we find the following two classes of solutions:
\begin{enumerate}
\item
For the first class of solutions, the subalgebra generated by the first three vector fields is mapped with the trivial homomorphism, and so \(\vec{j}_1 = \vec{j}_2 = \vec{j}_3 = \vec{k}_1 = \vec{k}_2 = \vec{k}_3 = 0\). The remaining element described by \(\vec{j}_4\) and \(\vec{k}_4\) is arbitrary. Hence, we find that there exists (up to equivalence by conjugacy classes) one algebra homomorphism for each orbit of the Lorentz group action on \(\mathfrak{so}(1,3)\) as described in section~\ref{sec:lorentz}.

\item
For the non-trivial case, we have the conditions~\eqref{eq:condII34}, \eqref{eq:condII4} and~\eqref{eq:condII3}. We start with the condition~\eqref{eq:condII4}, from which one can deduce \(\vec{j}_4 \neq 0\). We make use of the freedom of rotations in order to choose \(\vec{j}_4 \parallel \vec{e}_3\). Further, we must have \(\vec{j}_4 \perp \vec{k}_4\), and again we can rotate these vectors such that \(\vec{k}_4 \parallel \vec{e}_2\). From~\eqref{eq:condII4} finally also follows that the lengths of these two vectors are related; the general solution to this length condition is given by
\begin{equation}
\vec{j}_4 = \vec{e}_3\cosh u\,, \quad
\vec{k}_4 = \vec{e}_2\sinh u
\end{equation}
for some \(u \in \mathbb{R}\). From this form and the adjoint representation discussed in section~\ref{sec:lorentz} we see that we can perform a boost with \(K_1\), which acts additively on \(u\), and so we can achieve \(u = 0\). We thus choose \(\vec{j}_4 = \vec{e}_3\) and \(\vec{k}_4 = 0\). Proceeding with the condition~\eqref{eq:condII34}, we must have \(\vec{j}_3 \perp \vec{e}_3\) and \(\vec{k}_3 \perp \vec{e}_3\). Further, the condition~\eqref{eq:condII3} shows that \(\vec{j}_3 \perp \vec{k}_3\) as well as \(\|\vec{j}_3\| = \|\vec{k}_3\|\). By a suitable rotation using \(J_3\), we can thus always achieve \(\vec{j}_3 = c\vec{e}_2\) and \(\vec{k}_3 = c\vec{e}_1\) for some \(c > 0\), without interfering with the choice of \(\vec{j}_4\) which fixes one axis already. The same holds for a boost with \(K_3\), which allows us to perform a rescaling to set \(c = 1\). Having fixed these vectors, the conditions~\eqref{eq:vekII} then uniquely determine the remaining vectors as
\begin{equation}
\vec{j}_1 = \vec{k}_1 = 0\,, \quad
\vec{j}_2 = \vec{e}_1\,, \quad
\vec{k}_2 = -\vec{e}_2\,.
\end{equation}
Hence, we have found that all solutions within this branch are equivalent to each other up to Lorentz transformations, and we have found a representative for this unique equivalence class.
\end{enumerate}
We will make use of our findings in section~\ref{ssec:cosmoII} in order to derive the most general teleparallel geometry which is invariant under these symmetry generators.

\subsection{Bianchi type III}\label{ssec:algebraIII}
For the Bianchi type III, the three translation generators satisfy the relations
\begin{equation}\label{eq:algIII123}
[X_1, X_2] = [X_2, X_3] = 0\,, \quad
[X_1, X_3] = -X_1\,.
\end{equation}
As for the type II, the general homogeneous metric admits another symmetry under a vector field \(X_4\) which satisfies
\begin{equation}\label{eq:algIII4}
[X_1, X_4] = X_3\,, \quad
[X_2, X_4] = 0\,, \quad
[X_3, X_4] = -X_4\,.
\end{equation}
In terms of the cross product representation, a Lie algebra homomorphism to the Lorentz algebra must thus satisfy
\begin{subequations}\label{eq:vekIII}
\begin{align}
\vec{j}_1 \times \vec{j}_2 - \vec{k}_1 \times \vec{k}_2 &= 0\,, &
\vec{j}_1 \times \vec{k}_2 + \vec{k}_1 \times \vec{j}_2 &= 0\,,\label{eq:vekIIIa}\\
\vec{j}_1 \times \vec{j}_3 - \vec{k}_1 \times \vec{k}_3 &= -\vec{j}_1\,, &
\vec{j}_1 \times \vec{k}_3 + \vec{k}_1 \times \vec{j}_3 &= -\vec{k}_1\,,\label{eq:vekIIIb}\\
\vec{j}_1 \times \vec{j}_4 - \vec{k}_1 \times \vec{k}_4 &= \vec{j}_3\,, &
\vec{j}_1 \times \vec{k}_4 + \vec{k}_1 \times \vec{j}_4 &= \vec{k}_3\,,\label{eq:vekIIIc}\\
\vec{j}_2 \times \vec{j}_3 - \vec{k}_2 \times \vec{k}_3 &= 0\,, &
\vec{j}_2 \times \vec{k}_3 + \vec{k}_2 \times \vec{j}_3 &= 0\,,\label{eq:vekIIId}\\
\vec{j}_2 \times \vec{j}_4 - \vec{k}_2 \times \vec{k}_4 &= 0\,, &
\vec{j}_2 \times \vec{k}_4 + \vec{k}_2 \times \vec{j}_4 &= 0\,,\label{eq:vekIIIe}\\
\vec{j}_3 \times \vec{j}_4 - \vec{k}_3 \times \vec{k}_4 &= -\vec{j}_4\,, &
\vec{j}_3 \times \vec{k}_4 + \vec{k}_3 \times \vec{j}_4 &= -\vec{k}_4\,.\label{eq:vekIIIf}
\end{align}
\end{subequations}
We start with the third pair~\eqref{eq:vekIIIc} of equations, which we can immediately solve for \(\vec{j}_3\) and \(\vec{k}_3\), and then substitute into the remaining equations. In particular, this yields the equations
\begin{subequations}\label{eq:crossIII}
\begin{align}
\vec{j}_1 \times (\vec{j}_1 \times \vec{j}_4 - \vec{k}_1 \times \vec{k}_4) - \vec{k}_1 \times (\vec{j}_1 \times \vec{k}_4 + \vec{k}_1 \times \vec{j}_4) &= -\vec{j}_1\,,\label{eq:crossIIIa}\\
\vec{j}_1 \times (\vec{j}_1 \times \vec{k}_4 + \vec{k}_1 \times \vec{j}_4) + \vec{k}_1 \times (\vec{j}_1 \times \vec{j}_4 - \vec{k}_1 \times \vec{k}_4) &= -\vec{k}_1\,,\label{eq:crossIIIb}\\
\vec{j}_2 \times (\vec{j}_1 \times \vec{j}_4 - \vec{k}_1 \times \vec{k}_4) - \vec{k}_2 \times (\vec{j}_1 \times \vec{k}_4 + \vec{k}_1 \times \vec{j}_4) &= 0\,,\label{eq:crossIIIc}\\
\vec{j}_2 \times (\vec{j}_1 \times \vec{k}_4 + \vec{k}_1 \times \vec{j}_4) + \vec{k}_2 \times (\vec{j}_1 \times \vec{j}_4 - \vec{k}_1 \times \vec{k}_4) &= 0\,,\label{eq:crossIIId}\\
(\vec{j}_1 \times \vec{j}_4 - \vec{k}_1 \times \vec{k}_4) \times \vec{j}_4 - (\vec{j}_1 \times \vec{k}_4 + \vec{k}_1 \times \vec{j}_4) \times \vec{k}_4 &= -\vec{j}_4\,,\label{eq:crossIIIe}\\
(\vec{j}_1 \times \vec{j}_4 - \vec{k}_1 \times \vec{k}_4) \times \vec{k}_4 + (\vec{j}_1 \times \vec{k}_4 + \vec{k}_1 \times \vec{j}_4) \times \vec{j}_4 &= -\vec{k}_4\,.\label{eq:crossIIIf}
\end{align}
\end{subequations}
If we rewrite the double vector product in terms of scalar products, these take the form
\begin{subequations}\label{eq:dotIII}
\begin{align}
(\vec{j}_1 \cdot \vec{j}_1 - \vec{k}_1 \cdot \vec{k}_1)\vec{j}_4 - (\vec{j}_1 \cdot \vec{j}_4 - \vec{k}_1 \cdot \vec{k}_4)\vec{j}_1 + (\vec{j}_1 \cdot \vec{k}_4 + \vec{j}_4 \cdot \vec{k}_1)\vec{k}_1 - 2(\vec{j}_1 \cdot \vec{k}_1)\vec{k}_4 &= \vec{j}_1\,,\label{eq:dotIIIa}\\
(\vec{j}_1 \cdot \vec{j}_1 - \vec{k}_1 \cdot \vec{k}_1)\vec{k}_4 - (\vec{j}_1 \cdot \vec{j}_4 - \vec{k}_1 \cdot \vec{k}_4)\vec{k}_1 - (\vec{j}_1 \cdot \vec{k}_4 + \vec{j}_4 \cdot \vec{k}_1)\vec{j}_1 + 2(\vec{j}_1 \cdot \vec{k}_1)\vec{j}_4 &= \vec{k}_1\,,\label{eq:dotIIIb}\\
(\vec{j}_1 \cdot \vec{j}_2 - \vec{k}_1 \cdot \vec{k}_2)\vec{j}_4 - (\vec{j}_2 \cdot \vec{j}_4 - \vec{k}_2 \cdot \vec{k}_4)\vec{j}_1 + (\vec{j}_2 \cdot \vec{k}_4 + \vec{j}_4 \cdot \vec{k}_2)\vec{k}_1 - (\vec{j}_2 \cdot \vec{k}_1 + \vec{j}_1 \cdot \vec{k}_2)\vec{k}_4 &= 0\,,\label{eq:dotIIIc}\\
(\vec{j}_1 \cdot \vec{j}_2 - \vec{k}_1 \cdot \vec{k}_2)\vec{k}_4 - (\vec{j}_2 \cdot \vec{j}_4 - \vec{k}_2 \cdot \vec{k}_4)\vec{k}_1 - (\vec{j}_2 \cdot \vec{k}_4 + \vec{j}_4 \cdot \vec{k}_2)\vec{j}_1 + (\vec{j}_2 \cdot \vec{k}_1 + \vec{j}_1 \cdot \vec{k}_2)\vec{j}_4 &= 0\,,\label{eq:dotIIId}\\
(\vec{j}_4 \cdot \vec{j}_4 - \vec{k}_4 \cdot \vec{k}_4)\vec{j}_1 - (\vec{j}_4 \cdot \vec{j}_1 - \vec{k}_4 \cdot \vec{k}_1)\vec{j}_4 + (\vec{j}_4 \cdot \vec{k}_1 + \vec{j}_1 \cdot \vec{k}_4)\vec{k}_4 - 2(\vec{j}_4 \cdot \vec{k}_4)\vec{k}_1 &= \vec{j}_4\,,\label{eq:dotIIIe}\\
(\vec{j}_4 \cdot \vec{j}_4 - \vec{k}_4 \cdot \vec{k}_4)\vec{k}_1 - (\vec{j}_4 \cdot \vec{j}_1 - \vec{k}_4 \cdot \vec{k}_1)\vec{k}_4 - (\vec{j}_4 \cdot \vec{k}_1 + \vec{j}_1 \cdot \vec{k}_4)\vec{j}_4 + 2(\vec{j}_4 \cdot \vec{k}_4)\vec{j}_1 &= \vec{k}_4\,.\label{eq:dotIIIf}
\end{align}
\end{subequations}
We start with the first two equations~\eqref{eq:dotIIIa} and~\eqref{eq:dotIIIb}. Taking scalar products with \(\vec{j}_1\) and \(\vec{k}_1\), they become
\begin{subequations}\label{eq:ddotIII1}
\begin{align}
(\vec{j}_1 \cdot \vec{j}_1)(\vec{j}_4 \cdot \vec{k}_1) - (\vec{j}_1 \cdot \vec{j}_4)(\vec{j}_1 \cdot \vec{k}_1) + (\vec{j}_1 \cdot \vec{k}_4)(\vec{k}_1 \cdot \vec{k}_1) - (\vec{j}_1 \cdot \vec{k}_1)(\vec{k}_4 \cdot \vec{k}_1) &= \vec{j}_1 \cdot \vec{k}_1\,,\\
-(\vec{k}_1 \cdot \vec{k}_1)(\vec{k}_4 \cdot \vec{j}_1) + (\vec{k}_1 \cdot \vec{k}_4)(\vec{k}_1 \cdot \vec{j}_1) - (\vec{j}_4 \cdot \vec{k}_1)(\vec{j}_1 \cdot \vec{j}_1) + (\vec{j}_1 \cdot \vec{k}_1)(\vec{j}_4 \cdot \vec{j}_1) &= \vec{k}_1 \cdot \vec{j}_1\,,\\
-(\vec{k}_1 \cdot \vec{k}_1)(\vec{j}_4 \cdot \vec{j}_1) + (\vec{k}_1 \cdot \vec{k}_4)(\vec{j}_1 \cdot \vec{j}_1) + (\vec{j}_4 \cdot \vec{k}_1)(\vec{k}_1 \cdot \vec{j}_1) - (\vec{j}_1 \cdot \vec{k}_1)(\vec{k}_4 \cdot \vec{j}_1) &= \vec{j}_1 \cdot \vec{j}_1\,,\\
(\vec{j}_1 \cdot \vec{j}_1)(\vec{k}_4 \cdot \vec{k}_1) - (\vec{j}_1 \cdot \vec{j}_4)(\vec{k}_1 \cdot \vec{k}_1) - (\vec{j}_1 \cdot \vec{k}_4)(\vec{j}_1 \cdot \vec{k}_1) + (\vec{j}_1 \cdot \vec{k}_1)(\vec{j}_4 \cdot \vec{k}_1) &= \vec{k}_1 \cdot \vec{k}_1\,.
\end{align}
\end{subequations}
We see that we can take linear combinations such that their left hand sides cancel, and we end up with the equations
\begin{equation}\label{eq:condIII1}
\vec{j}_1 \cdot \vec{k}_1 = \vec{j}_1 \cdot \vec{j}_1 - \vec{k}_1 \cdot \vec{k}_1 = 0\,.
\end{equation}
We can proceed similarly with the last two equations~\eqref{eq:dotIIIe} and~\eqref{eq:dotIIIf}, where we now take scalar products with \(\vec{j}_4\) and \(\vec{k}_4\), thus obtaining
\begin{subequations}\label{eq:ddotIII4}
\begin{align}
(\vec{j}_4 \cdot \vec{j}_4)(\vec{j}_1 \cdot \vec{k}_4) - (\vec{j}_4 \cdot \vec{j}_1)(\vec{j}_4 \cdot \vec{k}_4) + (\vec{j}_4 \cdot \vec{k}_1)(\vec{k}_4 \cdot \vec{k}_4) - (\vec{j}_4 \cdot \vec{k}_4)(\vec{k}_1 \cdot \vec{k}_4) &= \vec{j}_4 \cdot \vec{k}_4\,,\\
-(\vec{k}_4 \cdot \vec{k}_4)(\vec{k}_1 \cdot \vec{j}_4) + (\vec{k}_4 \cdot \vec{k}_1)(\vec{k}_4 \cdot \vec{j}_4) - (\vec{j}_1 \cdot \vec{k}_4)(\vec{j}_4 \cdot \vec{j}_4) + (\vec{j}_4 \cdot \vec{k}_4)(\vec{j}_1 \cdot \vec{j}_4) &= \vec{k}_4 \cdot \vec{j}_4\,,\\
-(\vec{k}_4 \cdot \vec{k}_4)(\vec{j}_1 \cdot \vec{j}_4) + (\vec{k}_4 \cdot \vec{k}_1)(\vec{j}_4 \cdot \vec{j}_4) + (\vec{j}_1 \cdot \vec{k}_4)(\vec{k}_4 \cdot \vec{j}_4) - (\vec{j}_4 \cdot \vec{k}_4)(\vec{k}_1 \cdot \vec{j}_4) &= \vec{j}_4 \cdot \vec{j}_4\,,\\
(\vec{j}_4 \cdot \vec{j}_4)(\vec{k}_1 \cdot \vec{k}_4) - (\vec{j}_4 \cdot \vec{j}_1)(\vec{k}_4 \cdot \vec{k}_4) - (\vec{j}_4 \cdot \vec{k}_1)(\vec{j}_4 \cdot \vec{k}_4) + (\vec{j}_4 \cdot \vec{k}_4)(\vec{j}_1 \cdot \vec{k}_4) &= \vec{k}_4 \cdot \vec{k}_4\,.
\end{align}
\end{subequations}
Also here the left hand sides cancel for two suitable linear combinations, leaving us with the equations
\begin{equation}\label{eq:condIII4}
\vec{j}_4 \cdot \vec{k}_4 = \vec{j}_4 \cdot \vec{j}_4 - \vec{k}_4 \cdot \vec{k}_4 = 0\,.
\end{equation}
Substituting our findings back into the equations~\eqref{eq:dotIII} we started from, we now see that several terms vanish and they simplify to
\begin{subequations}\label{eq:dotIII14}
\begin{align}
-(\vec{j}_1 \cdot \vec{j}_4 - \vec{k}_1 \cdot \vec{k}_4)\vec{j}_1 + (\vec{j}_1 \cdot \vec{k}_4 + \vec{j}_4 \cdot \vec{k}_1)\vec{k}_1 &= \vec{j}_1\,,\label{eq:dotIII14a}\\
-(\vec{j}_1 \cdot \vec{j}_4 - \vec{k}_1 \cdot \vec{k}_4)\vec{k}_1 - (\vec{j}_1 \cdot \vec{k}_4 + \vec{j}_4 \cdot \vec{k}_1)\vec{j}_1 &= \vec{k}_1\,,\label{eq:dotIII14b}\\
-(\vec{j}_4 \cdot \vec{j}_1 - \vec{k}_4 \cdot \vec{k}_1)\vec{j}_4 + (\vec{j}_4 \cdot \vec{k}_1 + \vec{j}_1 \cdot \vec{k}_4)\vec{k}_4 &= \vec{j}_4\,,\label{eq:dotIII14c}\\
-(\vec{j}_4 \cdot \vec{j}_1 - \vec{k}_4 \cdot \vec{k}_1)\vec{k}_4 - (\vec{j}_4 \cdot \vec{k}_1 + \vec{j}_1 \cdot \vec{k}_4)\vec{j}_4 &= \vec{k}_4\,.\label{eq:dotIII14d}
\end{align}
\end{subequations}
It then follows that the scalar products~\eqref{eq:ddotIII1} and~\eqref{eq:ddotIII4} become
\begin{subequations}
\begin{align}
(\vec{j}_1 \cdot \vec{j}_1)(\vec{j}_1 \cdot \vec{k}_4 + \vec{j}_4 \cdot \vec{k}_1) &= 0\,, &
(\vec{j}_1 \cdot \vec{j}_1)(\vec{j}_1 \cdot \vec{j}_4 - \vec{k}_1 \cdot \vec{k}_4 + 1) &= 0\,,\\
(\vec{j}_4 \cdot \vec{j}_4)(\vec{j}_1 \cdot \vec{k}_4 + \vec{j}_4 \cdot \vec{k}_1) &= 0\,, &
(\vec{j}_4 \cdot \vec{j}_4)(\vec{j}_1 \cdot \vec{j}_4 - \vec{k}_1 \cdot \vec{k}_4 + 1) &= 0\,.
\end{align}
\end{subequations}
If we assume \(\vec{j}_1 \cdot \vec{k}_4 + \vec{j}_4 \cdot \vec{k}_1 \neq 0\), then the first equation in each row leads to \(\vec{j}_1 = \vec{j}_4 = 0\), and so also \(\vec{j}_1 \cdot \vec{k}_4 + \vec{j}_4 \cdot \vec{k}_1 = 0\), contradicting the assumption. Hence, we conclude that the latter is always satisfied. The remaining equations then lead to two distinct branches of solutions:
\begin{enumerate}
\item
In the case \(\vec{j}_1 = \vec{j}_4 = 0\), the previously found equations~\eqref{eq:condIII1} and~\eqref{eq:condIII4} imply that also \(\vec{k}_1 = \vec{k}_4 = 0\). In this case, from~\eqref{eq:vekIIIc} follows also \(\vec{j}_3 = \vec{k}_3 = 0\). With this, the remaining equations~\eqref{eq:vekIII} are satisfied identically, and \(\vec{j}_2\) and \(\vec{k}_2\) are arbitrary.

\item
Otherwise, we must have
\begin{equation}\label{eq:condIII14}
\vec{j}_1 \cdot \vec{k}_4 + \vec{j}_4 \cdot \vec{k}_1 = \vec{j}_1 \cdot \vec{j}_4 - \vec{k}_1 \cdot \vec{k}_4 + 1 = 0\,,
\end{equation}
where the latter in particular implies that the homomorphism is non-trivial. Substituting these conditions in~\eqref{eq:dotIII14}, we see that these equations are now satisfied identically.
\end{enumerate}
For the second branch, we are still left with the equations involving \(\vec{j}_2\) and \(\vec{k}_2\). These are the original equations~\eqref{eq:vekIIIa} and~\eqref{eq:vekIIIe}, as well as~\eqref{eq:vekIIId}, which yields~\eqref{eq:dotIIIc} and~\eqref{eq:dotIIId} after substituting \(\vec{j}_3\) and \(\vec{k}_3\) from~\eqref{eq:vekIIIc}. It is most helpful to start with the latter equations, and take their scalar product with \(\vec{j}_1, \vec{j}_4, \vec{k}_1, \vec{k}_4\), where we use the relations~\eqref{eq:condIII1}, \eqref{eq:condIII4} and~\eqref{eq:condIII14} to cancel numerous terms. One then finds
\begin{subequations}
\begin{align}
(\vec{j}_1 \cdot \vec{j}_2 - \vec{k}_1 \cdot \vec{k}_2)(\vec{j}_1 \cdot \vec{j}_4) - (\vec{j}_2 \cdot \vec{j}_4 - \vec{k}_2 \cdot \vec{k}_4)(\vec{j}_1 \cdot \vec{j}_1) - (\vec{j}_2 \cdot \vec{k}_1 + \vec{j}_1 \cdot \vec{k}_2)(\vec{j}_1 \cdot \vec{k}_4) &= 0\,,\\
(\vec{j}_1 \cdot \vec{j}_2 - \vec{k}_1 \cdot \vec{k}_2)(\vec{j}_1 \cdot \vec{k}_4) - (\vec{j}_2 \cdot \vec{k}_4 + \vec{j}_4 \cdot \vec{k}_2)(\vec{j}_1 \cdot \vec{j}_1) + (\vec{j}_2 \cdot \vec{k}_1 + \vec{j}_1 \cdot \vec{k}_2)(\vec{j}_1 \cdot \vec{j}_4) &= 0\,,\\
(\vec{j}_1 \cdot \vec{j}_2 - \vec{k}_1 \cdot \vec{k}_2)(\vec{j}_4 \cdot \vec{j}_4) - (\vec{j}_2 \cdot \vec{j}_4 - \vec{k}_2 \cdot \vec{k}_4)(\vec{j}_1 \cdot \vec{j}_4) + (\vec{j}_2 \cdot \vec{k}_4 - \vec{j}_4 \cdot \vec{k}_2)(\vec{j}_1 \cdot \vec{k}_4) &= 0\,,\\
(\vec{j}_2 \cdot \vec{j}_4 - \vec{k}_2 \cdot \vec{k}_4)(\vec{j}_1 \cdot \vec{k}_4) - (\vec{j}_2 \cdot \vec{k}_4 + \vec{j}_4 \cdot \vec{k}_2)(\vec{j}_1 \cdot \vec{j}_4) + (\vec{j}_2 \cdot \vec{k}_1 + \vec{j}_1 \cdot \vec{k}_2)(\vec{j}_4 \cdot \vec{j}_4) &= 0\,,\\
-(\vec{j}_1 \cdot \vec{j}_2 - \vec{k}_1 \cdot \vec{k}_2)(\vec{j}_1 \cdot \vec{k}_4) + (\vec{j}_2 \cdot \vec{k}_4 + \vec{j}_4 \cdot \vec{k}_2)(\vec{j}_1 \cdot \vec{j}_1) - (\vec{j}_2 \cdot \vec{k}_1 + \vec{j}_1 \cdot \vec{k}_2)(\vec{j}_1 \cdot \vec{j}_4 + 1) &= 0\,,\\
(\vec{j}_1 \cdot \vec{j}_2 - \vec{k}_1 \cdot \vec{k}_2)(\vec{j}_1 \cdot \vec{j}_4 + 1) - (\vec{j}_2 \cdot \vec{j}_4 - \vec{k}_2 \cdot \vec{k}_4)(\vec{j}_1 \cdot \vec{j}_1) - (\vec{j}_2 \cdot \vec{k}_1 + \vec{j}_1 \cdot \vec{k}_2)(\vec{j}_1 \cdot \vec{k}_4) &= 0\,,\\
-(\vec{j}_2 \cdot \vec{j}_4 - \vec{k}_2 \cdot \vec{k}_4)(\vec{j}_1 \cdot \vec{k}_4) + (\vec{j}_2 \cdot \vec{k}_4 + \vec{j}_4 \cdot \vec{k}_2)(\vec{j}_1 \cdot \vec{j}_4 + 1) - (\vec{j}_2 \cdot \vec{k}_1 + \vec{j}_1 \cdot \vec{k}_2)(\vec{j}_4 \cdot \vec{j}_4) &= 0\,,\\
(\vec{j}_1 \cdot \vec{j}_2 - \vec{k}_1 \cdot \vec{k}_2)(\vec{j}_4 \cdot \vec{j}_4) - (\vec{j}_2 \cdot \vec{j}_4 - \vec{k}_2 \cdot \vec{k}_4)(\vec{j}_1 \cdot \vec{j}_4 + 1) - (\vec{j}_2 \cdot \vec{k}_4 + \vec{j}_4 \cdot \vec{k}_2)(\vec{j}_1 \cdot \vec{k}_4) &= 0\,,
\end{align}
\end{subequations}
Now we see that for each of the last four equations there is one of the first four equations such that all terms except for a single quadratic term cancel, leaving the equations
\begin{equation}
\vec{j}_1 \cdot \vec{j}_2 - \vec{k}_1 \cdot \vec{k}_2 = \vec{j}_2 \cdot \vec{j}_4 - \vec{k}_2 \cdot \vec{k}_4 = \vec{j}_2 \cdot \vec{k}_4 + \vec{j}_4 \cdot \vec{k}_2 = \vec{j}_2 \cdot \vec{k}_1 + \vec{j}_1 \cdot \vec{k}_2 = 0\,.
\end{equation}
Substituting these conditions into the equations~\eqref{eq:dotIIIc} and~\eqref{eq:dotIIId}, we see that these are now solved, and we are only left with the algebra relations~\eqref{eq:vekIIIa} and~\eqref{eq:vekIIIe}. We can substitute the previously found relations and use the first one and calculate
\begin{subequations}
\begin{align}
0 &= \vec{j}_4 \times (\vec{j}_1 \times \vec{j}_2 - \vec{k}_1 \times \vec{k}_2) - \vec{k}_4 \times (\vec{j}_1 \times \vec{k}_2 - \vec{j}_2 \times \vec{k}_1) = \vec{j}_2\,,\\
0 &= \vec{k}_4 \times (\vec{j}_1 \times \vec{j}_2 - \vec{k}_1 \times \vec{k}_2) + \vec{j}_4 \times (\vec{j}_1 \times \vec{k}_2 - \vec{j}_2 \times \vec{k}_1) = \vec{k}_2\,,
\end{align}
\end{subequations}
showing that these two vectors must vanish. One easily checks that this solves all remaining equations. In summary, we thus have the following two solutions:
\begin{enumerate}
\item
In this case, the subalgebra generated by the vector fields \(X_1, X_3, X_4\) is mapped with the trivial homomorphism, and so \(\vec{j}_1 = \vec{j}_3 = \vec{j}_4 = \vec{k}_1 = \vec{k}_3 = \vec{k}_4 = 0\). The element \(\boldsymbol{\lambda}_2\) described by \(\vec{j}_2\) and \(\vec{k}_2\) is arbitrary. We thus find that the inequivalent Lie algebra homomorphisms are in one-to-one correspondence with the orbits of the Lorentz group action on \(\mathfrak{so}(1,3)\) as described in section~\ref{sec:lorentz}.

\item
Considering the non-trivial case, the condition~\eqref{eq:condIII1} implies \(\vec{j}_1 \perp \vec{k}_1\) and \(\|\vec{j}_1\| = \|\vec{k}_1\|\). By a suitable rotation, we can thus set \(\vec{j}_1 = c\vec{e}_1\) and \(\vec{k}_1 = c\vec{e}_2\) for some \(c > 0\). Using a boost with \(K_3\), which acts as a rescaling of these two vectors with a common positive scale factor, we can set \(c\) to any desired value, and will do so later to normalize the solution. By the same argument drawn from~\eqref{eq:condIII4}, we have \(\vec{j}_4 \perp \vec{k}_4\) and \(\|\vec{j}_4\| = \|\vec{k}_4\|\). Further, from~\eqref{eq:condIII14} follows
\begin{equation}
\vec{e}_3 \cdot \vec{k}_4 + \vec{j}_4 \cdot \vec{e}_1 = \vec{e}_3 \cdot \vec{j}_4 - \vec{e}_1 \cdot \vec{k}_4 + c^{-1} = 0\,.
\end{equation}
There exists a two-parameter family of solutions to these equations, given by
\begin{equation}
\vec{j}_4 = \frac{(a^2 - b^2)c^2 - 1}{2c}\vec{e}_1 - abc\vec{e}_2 + a\vec{e}_3\,, \quad
\vec{k}_4 = abc\vec{e}_1 + \frac{(a^2 - b^2)c^2 + 1}{2c}\vec{e}_2 + b\vec{e}_3\,,
\end{equation}
with \(a, b \in \mathbb{R}\). We can still simplify this solution by realizing that there exists a two-parameter subgroup of the Lorentz group which leaves \(\vec{j}_1\) and \(\vec{k}_1\) invariant, which is generated by the elements \(J_1 + K_2\) and \(J_2 - K_1\). By applying the transformation
\begin{equation}
\exp(-ac(J_2 - K_1) - bc(J_1 + K_2))\,,
\end{equation}
one obtains
\begin{equation}
\vec{j}_4 = -\frac{1}{2c}\vec{e}_1\,, \quad
\vec{k}_4 = \frac{1}{2c}\vec{e}_2\,.
\end{equation}
Now we can set \(c = 1/\sqrt{2}\) as a normalization. From the relations~\eqref{eq:vekIII} then finally follows
\begin{equation}
\vec{j}_2 = \vec{j}_3 = \vec{k}_2 = 0\,, \quad
\vec{k}_3 = \vec{e}_3\,.
\end{equation}
Hence, we find a unique solution up to Lorentz transformations.
\end{enumerate}
We will derive the corresponding teleparallel geometry in section~\ref{ssec:cosmoIII}.

\subsection{Bianchi type IX}\label{ssec:algebraIX}
For the Bianchi type IX, the three translation generators satisfy the relations
\begin{equation}\label{eq:algIX123}
[X_1, X_2] = X_3\,, \quad
[X_2, X_3] = X_1\,, \quad
[X_3, X_1] = X_2\,,
\end{equation}
and so one recognizes that their Lie algebra is given by \(\mathfrak{so}(3)\). One finds that the general homogeneous metric admits another symmetry under a vector field \(X_4\) which commutes with the translation generators,
\begin{equation}\label{eq:algIX4}
[X_1, X_4] = [X_2, X_4] = [X_3, X_4] = 0\,.
\end{equation}
The Lie algebra homomorphisms \(\boldsymbol{\lambda}: \mathfrak{so}(3) \to \mathfrak{so}(1,3)\) are, of course, well-known, and this has been used to derive teleparallel geometries in~\cite{Hohmann:2019nat}. It is nevertheless instructive to follow the procedure we have developed also in this case. From the algebra relations~\eqref{eq:algIX123} follows
\begin{subequations}\label{eq:vekIX123}
\begin{align}
\vec{j}_2 \times \vec{j}_3 - \vec{k}_2 \times \vec{k}_3 &= \vec{j}_1\,, &
\vec{j}_2 \times \vec{k}_3 + \vec{k}_2 \times \vec{j}_3 &= \vec{k}_1\,,\\
\vec{j}_3 \times \vec{j}_1 - \vec{k}_3 \times \vec{k}_1 &= \vec{j}_2\,, &
\vec{j}_3 \times \vec{k}_1 + \vec{k}_3 \times \vec{j}_1 &= \vec{k}_2\,,\\
\vec{j}_1 \times \vec{j}_2 - \vec{k}_1 \times \vec{k}_2 &= \vec{j}_3\,, &
\vec{j}_1 \times \vec{k}_2 + \vec{k}_1 \times \vec{j}_2 &= \vec{k}_3\,.
\end{align}
\end{subequations}
The first line can be used to solve for \(\vec{j}_1\) and \(\vec{k}_1\). Inserting into the remaining equations then yields
\begin{subequations}\label{eq:crossIX23}
\begin{align}
\vec{j}_3 \times (\vec{j}_2 \times \vec{j}_3 - \vec{k}_2 \times \vec{k}_3) - \vec{k}_3 \times (\vec{j}_2 \times \vec{k}_3 + \vec{k}_2 \times \vec{j}_3) &= \vec{j}_2\,,\\
\vec{j}_3 \times (\vec{j}_2 \times \vec{k}_3 + \vec{k}_2 \times \vec{j}_3) + \vec{k}_3 \times (\vec{j}_2 \times \vec{j}_3 - \vec{k}_2 \times \vec{k}_3) &= \vec{k}_2\,,\\
(\vec{j}_2 \times \vec{j}_3 - \vec{k}_2 \times \vec{k}_3) \times \vec{j}_2 - (\vec{j}_2 \times \vec{k}_3 + \vec{k}_2 \times \vec{j}_3) \times \vec{k}_2 &= \vec{j}_3\,,\\
(\vec{j}_2 \times \vec{j}_3 - \vec{k}_2 \times \vec{k}_3) \times \vec{k}_2 + (\vec{j}_2 \times \vec{k}_3 + \vec{k}_2 \times \vec{j}_3) \times \vec{j}_2 &= \vec{k}_3\,.
\end{align}
\end{subequations}
Rewriting the double vector product, they become
\begin{subequations}\label{eq:dotIX23}
\begin{align}
(\vec{j}_3 \cdot \vec{j}_3 - \vec{k}_3 \cdot \vec{k}_3)\vec{j}_2 - (\vec{j}_2 \cdot \vec{j}_3 - \vec{k}_2 \cdot \vec{k}_3)\vec{j}_3 - 2(\vec{j}_3 \cdot \vec{k}_3)\vec{k}_2 + (\vec{j}_2 \cdot \vec{k}_3 + \vec{j}_3 \cdot \vec{k}_2)\vec{k}_3 &= \vec{j}_2\,,\label{eq:dotIX23a}\\
(\vec{j}_3 \cdot \vec{j}_3 - \vec{k}_3 \cdot \vec{k}_3)\vec{k}_2 - (\vec{j}_2 \cdot \vec{j}_3 - \vec{k}_2 \cdot \vec{k}_3)\vec{k}_3 + 2(\vec{j}_3 \cdot \vec{k}_3)\vec{j}_2 - (\vec{j}_2 \cdot \vec{k}_3 + \vec{j}_3 \cdot \vec{k}_2)\vec{j}_3 &= \vec{k}_2\,,\label{eq:dotIX23b}\\
(\vec{j}_2 \cdot \vec{j}_2 - \vec{k}_2 \cdot \vec{k}_2)\vec{j}_3 - (\vec{j}_2 \cdot \vec{j}_3 - \vec{k}_2 \cdot \vec{k}_3)\vec{j}_2 - 2(\vec{j}_2 \cdot \vec{k}_2)\vec{k}_3 + (\vec{j}_2 \cdot \vec{k}_3 + \vec{j}_3 \cdot \vec{k}_2)\vec{k}_2 &= \vec{j}_3\,,\label{eq:dotIX23c}\\
(\vec{j}_2 \cdot \vec{j}_2 - \vec{k}_2 \cdot \vec{k}_2)\vec{k}_3 - (\vec{j}_2 \cdot \vec{j}_3 - \vec{k}_2 \cdot \vec{k}_3)\vec{k}_2 + 2(\vec{j}_2 \cdot \vec{k}_2)\vec{j}_3 - (\vec{j}_2 \cdot \vec{k}_3 + \vec{j}_3 \cdot \vec{k}_2)\vec{j}_2 &= \vec{k}_3\,.\label{eq:dotIX23d}
\end{align}
\end{subequations}
We then take the scalar product of the first equation~\eqref{eq:dotIX23a} with \(\vec{k}_3\), which reads
\begin{equation}
(\vec{j}_3 \cdot \vec{j}_3)(\vec{j}_2 \cdot \vec{k}_3) - (\vec{j}_2 \cdot \vec{j}_3)(\vec{j}_3 \cdot \vec{k}_3) - (\vec{j}_3 \cdot \vec{k}_3)(\vec{k}_2 \cdot \vec{k}_3) + (\vec{j}_3 \cdot \vec{k}_2)(\vec{k}_3 \cdot \vec{k}_3) = \vec{j}_2 \cdot \vec{k}_3\,,
\end{equation}
where several terms have canceled. Similarly, we multiply the second equation~\eqref{eq:dotIX23b} by \(\vec{j}_3\) and find
\begin{equation}
-(\vec{k}_3 \cdot \vec{k}_3)(\vec{k}_2 \cdot \vec{j}_3) + (\vec{k}_2 \cdot \vec{k}_3)(\vec{k}_3 \cdot \vec{j}_3) + (\vec{j}_3 \cdot \vec{k}_3)(\vec{j}_2 \cdot \vec{j}_3) - (\vec{j}_2 \cdot \vec{k}_3)(\vec{j}_3 \cdot \vec{j}_3) = \vec{k}_2 \cdot \vec{j}_3\,.
\end{equation}
Now we see that the left hand sides of these equations agree up to their sign, and so by adding them we obtain the relation
\begin{equation}\label{eq:j2k3j3k3IX}
\vec{j}_2 \cdot \vec{k}_3 + \vec{j}_3 \cdot \vec{k}_2 = 0\,.
\end{equation}
We can obtain another, similarly simple equation by taking the scalar product of the first equation~\eqref{eq:dotIX23a} with \(\vec{j}_3\), which becomes
\begin{equation}
-(\vec{k}_3 \cdot \vec{k}_3)(\vec{j}_2 \cdot \vec{j}_3) + (\vec{k}_2 \cdot \vec{k}_3)(\vec{j}_3 \cdot \vec{j}_3) - (\vec{j}_3 \cdot \vec{k}_3)(\vec{k}_2 \cdot \vec{j}_3) + (\vec{j}_2 \cdot \vec{k}_3)(\vec{k}_3 \cdot \vec{j}_3) = \vec{j}_2 \cdot \vec{j}_3\,,
\end{equation}
as well as multiplying the second equation~\eqref{eq:dotIX23b} by \(\vec{k}_3\), which reads
\begin{equation}
(\vec{j}_3 \cdot \vec{j}_3)(\vec{k}_2 \cdot \vec{k}_3) - (\vec{j}_2 \cdot \vec{j}_3)(\vec{k}_3 \cdot \vec{k}_3) + (\vec{j}_3 \cdot \vec{k}_3)(\vec{j}_2 \cdot \vec{k}_3) - (\vec{j}_3 \cdot \vec{k}_2)(\vec{j}_3 \cdot \vec{k}_3) = \vec{k}_2 \cdot \vec{k}_3\,,
\end{equation}
and taking their difference. Then the left hand sides cancel and we find
\begin{equation}\label{eq:j2j3k2k3IX}
\vec{j}_2 \cdot \vec{j}_3 - \vec{k}_2 \cdot \vec{k}_3 = 0\,.
\end{equation}
Note that we could have obtained the same relations by using the equations~\eqref{eq:dotIX23c} and~\eqref{eq:dotIX23d} instead. Inserting these into the previously derived equations~\eqref{eq:dotIX23}, we see that numerous terms cancel and we find
\begin{subequations}\label{eq:dotxIX23}
\begin{align}
(\vec{j}_3 \cdot \vec{j}_3 - \vec{k}_3 \cdot \vec{k}_3)\vec{j}_2 - 2(\vec{j}_3 \cdot \vec{k}_3)\vec{k}_2 &= \vec{j}_2\,,\label{eq:dotxIX23a}\\
(\vec{j}_3 \cdot \vec{j}_3 - \vec{k}_3 \cdot \vec{k}_3)\vec{k}_2 + 2(\vec{j}_3 \cdot \vec{k}_3)\vec{j}_2 &= \vec{k}_2\,,\label{eq:dotxIX23b}\\
(\vec{j}_2 \cdot \vec{j}_2 - \vec{k}_2 \cdot \vec{k}_2)\vec{j}_3 - 2(\vec{j}_2 \cdot \vec{k}_2)\vec{k}_3 &= \vec{j}_3\,,\label{eq:dotxIX23c}\\
(\vec{j}_2 \cdot \vec{j}_2 - \vec{k}_2 \cdot \vec{k}_2)\vec{k}_3 + 2(\vec{j}_2 \cdot \vec{k}_2)\vec{j}_3 &= \vec{k}_3\,.\label{eq:dotxIX23d}
\end{align}
\end{subequations}
Now we calculate the scalar products
\begin{subequations}
\begin{align}
(\vec{j}_3 \cdot \vec{j}_3 - \vec{k}_3 \cdot \vec{k}_3)(\vec{j}_2 \cdot \vec{k}_2) - 2(\vec{j}_3 \cdot \vec{k}_3)(\vec{k}_2 \cdot \vec{k}_2) &= \vec{j}_2 \cdot \vec{k}_2\,,\\
(\vec{j}_3 \cdot \vec{j}_3 - \vec{k}_3 \cdot \vec{k}_3)(\vec{k}_2 \cdot \vec{j}_2) + 2(\vec{j}_3 \cdot \vec{k}_3)(\vec{j}_2 \cdot \vec{j}_2) &= \vec{k}_2 \cdot \vec{j}_2\,,\\
(\vec{j}_2 \cdot \vec{j}_2 - \vec{k}_2 \cdot \vec{k}_2)(\vec{j}_3 \cdot \vec{k}_3) - 2(\vec{j}_2 \cdot \vec{k}_2)(\vec{k}_3 \cdot \vec{k}_3) &= \vec{j}_3 \cdot \vec{k}_3\,,\\
(\vec{j}_2 \cdot \vec{j}_2 - \vec{k}_2 \cdot \vec{k}_2)(\vec{k}_3 \cdot \vec{j}_3) + 2(\vec{j}_2 \cdot \vec{k}_2)(\vec{j}_3 \cdot \vec{j}_3) &= \vec{k}_3 \cdot \vec{j}_3\,.
\end{align}
\end{subequations}
The difference of the first two equations as well as the difference of the last two equations then become
\begin{subequations}
\begin{align}
2(\vec{j}_3 \cdot \vec{k}_3)(\vec{j}_2 \cdot \vec{j}_2 + \vec{k}_2 \cdot \vec{k}_2) &= 0\,,\label{eq:s2322IX}\\
2(\vec{j}_2 \cdot \vec{k}_2)(\vec{j}_3 \cdot \vec{j}_3 + \vec{k}_3 \cdot \vec{k}_3) &= 0\,,\label{eq:s2333IX}
\end{align}
\end{subequations}
Further calculating the scalar products
\begin{subequations}
\begin{align}
(\vec{j}_3 \cdot \vec{j}_3 - \vec{k}_3 \cdot \vec{k}_3)(\vec{j}_2 \cdot \vec{j}_2) - 2(\vec{j}_3 \cdot \vec{k}_3)(\vec{k}_2 \cdot \vec{j}_2) &= \vec{j}_2 \cdot \vec{j}_2\,,\\
(\vec{j}_3 \cdot \vec{j}_3 - \vec{k}_3 \cdot \vec{k}_3)(\vec{k}_2 \cdot \vec{k}_2) + 2(\vec{j}_3 \cdot \vec{k}_3)(\vec{j}_2 \cdot \vec{k}_2) &= \vec{k}_2 \cdot \vec{k}_2\,,\\
(\vec{j}_2 \cdot \vec{j}_2 - \vec{k}_2 \cdot \vec{k}_2)(\vec{j}_3 \cdot \vec{j}_3) - 2(\vec{j}_2 \cdot \vec{k}_2)(\vec{k}_3 \cdot \vec{j}_3) &= \vec{j}_3 \cdot \vec{j}_3\,,\\
(\vec{j}_2 \cdot \vec{j}_2 - \vec{k}_2 \cdot \vec{k}_2)(\vec{k}_3 \cdot \vec{k}_3) + 2(\vec{j}_2 \cdot \vec{k}_2)(\vec{j}_3 \cdot \vec{k}_3) &= \vec{k}_3 \cdot \vec{k}_3\,,
\end{align}
\end{subequations}
the sums of the first two equations and the last two equations become, respectively,
\begin{subequations}
\begin{align}
(\vec{j}_3 \cdot \vec{j}_3 - \vec{k}_3 \cdot \vec{k}_3 - 1)(\vec{j}_2 \cdot \vec{j}_2 + \vec{k}_2 \cdot \vec{k}_2) &= 0\,\label{eq:n3322IX}\\
(\vec{j}_2 \cdot \vec{j}_2 - \vec{k}_2 \cdot \vec{k}_2 - 1)(\vec{j}_3 \cdot \vec{j}_3 + \vec{k}_3 \cdot \vec{k}_3) &= 0\,\label{eq:n2233IX}
\end{align}
\end{subequations}
We find that there are two branches of solutions to these equations:
\begin{enumerate}
\item
If we solve the equations~\eqref{eq:s2322IX} and~\eqref{eq:n3322IX} by \(\vec{j}_2 \cdot \vec{j}_2 + \vec{k}_2 \cdot \vec{k}_2 = 0\), then it follows from the fact that neither of these two terms can be negative that they must vanish individually, \(\vec{j}_2 \cdot \vec{j}_2 = \vec{k}_2 \cdot \vec{k}_2 = 0\), hence \(\vec{j}_2 = \vec{k}_2 = 0\). From the condition~\eqref{eq:n2233IX} then follows analogously that also \(\vec{j}_3 = \vec{k}_3 = 0\). From the original relation~\eqref{eq:crossIX23} then follows \(\vec{j}_1 = \vec{k}_1 = 0\). This is the \emph{trivial homomorphism}.

\item
Alternatively, we can solve the equations by setting
\begin{equation}
\vec{j}_2 \cdot \vec{k}_2 = \vec{j}_3 \cdot \vec{k}_3 = \vec{j}_2 \cdot \vec{j}_2 - \vec{k}_2 \cdot \vec{k}_2 - 1 = \vec{j}_3 \cdot \vec{j}_3 - \vec{k}_3 \cdot \vec{k}_3 - 1 = 0\,.\label{eq:nt23IX}
\end{equation}
Inserting these conditions in the equations~\eqref{eq:dotxIX23}, we find that these are now solved. Hence, together with the conditions~\eqref{eq:j2k3j3k3IX} and~\eqref{eq:j2j3k2k3IX} they constitute another full class of solutions to the original system of equations.
\end{enumerate}
Having obtained a set of simple both necessary and sufficient conditions for the first three vector fields, we finally come to the relations~\eqref{eq:algIX4} involving \(X_4\). These imply the homomorphism conditions
\begin{subequations}\label{eq:vekIX4}
\begin{align}
\vec{j}_1 \times \vec{j}_4 - \vec{k}_1 \times \vec{k}_4 = \vec{j}_2 \times \vec{j}_4 - \vec{k}_2 \times \vec{k}_4 = \vec{j}_3 \times \vec{j}_4 - \vec{k}_3 \times \vec{k}_4 &= 0\,,\\
\vec{j}_1 \times \vec{k}_4 + \vec{k}_1 \times \vec{j}_4 = \vec{j}_2 \times \vec{k}_4 + \vec{k}_2 \times \vec{j}_4 = \vec{j}_3 \times \vec{k}_4 + \vec{k}_3 \times \vec{j}_4 &= 0\,.
\end{align}
\end{subequations}
For the trivial homomorphism \(\boldsymbol{\lambda}_{1,2,3} = 0\) these are obviously satisfied for arbitrary \(\boldsymbol{\lambda}_4\), and so we are left with the non-trivial case. If we insert \(\vec{j}_1\) and \(\vec{k}_1\) obtained from the relations~\eqref{eq:vekIX123}, we can use the Jacobi identity of the vector product to obtain
\begin{equation}
\begin{split}
\vec{j}_1 \times \vec{j}_4 - \vec{k}_1 \times \vec{k}_4 &= (\vec{j}_2 \times \vec{j}_3 - \vec{k}_2 \times \vec{k}_3) \times \vec{j}_4 - (\vec{j}_2 \times \vec{k}_3 + \vec{k}_2 \times \vec{j}_3) \times \vec{k}_4\\
&= (\vec{j}_2 \times \vec{j}_4 - \vec{k}_2 \times \vec{k}_4) \times \vec{j}_3 - (\vec{j}_3 \times \vec{j}_4 - \vec{k}_3 \times \vec{k}_4) \times \vec{j}_2\\
&\phantom{=}+ (\vec{k}_3 \times \vec{j}_4 + \vec{j}_3 \times \vec{k}_4) \times \vec{k}_2 - (\vec{k}_2 \times \vec{j}_4 + \vec{j}_2 \times \vec{k}_4) \times \vec{k}_3
\end{split}
\end{equation}
and
\begin{equation}
\begin{split}
\vec{j}_1 \times \vec{k}_4 + \vec{k}_1 \times \vec{j}_4 &= (\vec{j}_2 \times \vec{j}_3 - \vec{k}_2 \times \vec{k}_3) \times \vec{k}_4 - (\vec{j}_2 \times \vec{k}_3 + \vec{k}_2 \times \vec{j}_3) \times \vec{j}_4\\
&= (\vec{j}_2 \times \vec{k}_4 + \vec{k}_2 \times \vec{j}_4) \times \vec{j}_3 - (\vec{j}_3 \times \vec{k}_4 + \vec{k}_3 \times \vec{j}_4) \times \vec{j}_2\\
&\phantom{=}+ (\vec{k}_3 \times \vec{k}_4 - \vec{j}_3 \times \vec{j}_4) \times \vec{k}_2 - (\vec{k}_2 \times \vec{k}_4 - \vec{j}_2 \times \vec{j}_4) \times \vec{k}_3\,,
\end{split}
\end{equation}
showing that these are satisfied if the remaining conditions on \(\boldsymbol{\lambda}_4\) are satisfied; this is of course an immediate consequence of the Jacobi identity
\begin{equation}
[X_1, X_4] = [[X_2, X_3], X_4] = -[[X_3, X_4], X_2] - [[X_4, X_2], X_3]\,.
\end{equation}
We can use similar equations in order to derive simpler necessary, and ultimately also sufficient, conditions on \(\boldsymbol{\lambda}_4\), by rewriting the double vector product as a scalar product, and forming further scalar products thereof. In particular, we find the relations
\begin{subequations}
\begin{align}
[(\vec{j}_2 \times \vec{k}_4 - \vec{j}_4 \times \vec{k}_2) \times \vec{j}_3 + (\vec{j}_2 \times \vec{j}_4 - \vec{k}_2 \times \vec{k}_4) \times \vec{k}_3] \cdot \vec{j}_2 &= -(\vec{j}_2 \cdot \vec{j}_2)(\vec{j}_3 \cdot \vec{k}_4 + \vec{j}_4 \cdot \vec{k}_3)\,,\\
[(\vec{j}_3 \times \vec{k}_4 - \vec{j}_4 \times \vec{k}_3) \times \vec{j}_2 + (\vec{j}_3 \times \vec{j}_4 - \vec{k}_3 \times \vec{k}_4) \times \vec{k}_2] \cdot \vec{j}_3 &= -(\vec{j}_3 \cdot \vec{j}_3)(\vec{j}_2 \cdot \vec{k}_4 + \vec{j}_4 \cdot \vec{k}_2)\,,\\
[(\vec{j}_2 \times \vec{k}_4 - \vec{j}_4 \times \vec{k}_2) \times \vec{k}_3 + (\vec{j}_2 \times \vec{j}_4 - \vec{k}_2 \times \vec{k}_4) \times \vec{j}_3] \cdot \vec{j}_2 &= (\vec{j}_2 \cdot \vec{j}_2)(\vec{j}_3 \cdot \vec{j}_4 - \vec{k}_3 \cdot \vec{k}_4)\,,\\
[(\vec{j}_3 \times \vec{k}_4 - \vec{j}_4 \times \vec{k}_3) \times \vec{k}_2 + (\vec{j}_3 \times \vec{j}_4 - \vec{k}_3 \times \vec{k}_4) \times \vec{j}_2] \cdot \vec{j}_3 &= (\vec{j}_3 \cdot \vec{j}_3)(\vec{j}_2 \cdot \vec{j}_4 - \vec{k}_2 \cdot \vec{k}_4)\,,
\end{align}
\end{subequations}
where we have made use of the conditions on \(\boldsymbol{\lambda}_{2,3}\) derived earlier to cancel several terms. From the algebra relations~\eqref{eq:vekIX4} follows that the left hand side vanishes, and so must also the right hand side. Further, note that from~\eqref{eq:nt23IX} in particular follows
\begin{equation}
\vec{j}_2 \cdot \vec{j}_2 = \vec{k}_2 \cdot \vec{k}_2 + 1 \geq 1\,, \quad
\vec{j}_3 \cdot \vec{j}_3 = \vec{k}_3 \cdot \vec{k}_3 + 1 \geq 1\,,
\end{equation}
and so these expressions must be non-vanishing. One therefore reads off the conditions
\begin{equation}
\vec{j}_3 \cdot \vec{k}_4 + \vec{j}_4 \cdot \vec{k}_3 = \vec{j}_2 \cdot \vec{k}_4 + \vec{j}_4 \cdot \vec{k}_2 = \vec{j}_3 \cdot \vec{j}_4 - \vec{k}_3 \cdot \vec{k}_4 = \vec{j}_2 \cdot \vec{j}_4 - \vec{k}_2 \cdot \vec{k}_4 = 0\,.
\end{equation}
We can therefore use these relations and, e.g., substitute the second term of each of the vanishing expressions by (the negative of) the first term in subsequent calculations. With these substitutions at hand, we find
\begin{equation}
[(\vec{j}_2 \times \vec{k}_4 - \vec{j}_4 \times \vec{k}_2) \times \vec{j}_2] \cdot \vec{j}_4 = (\vec{j}_2 \cdot \vec{j}_2)(\vec{j}_4 \cdot \vec{k}_4)\,,
\end{equation}
leading to the conclusion
\begin{equation}
\vec{j}_4 \cdot \vec{k}_4 = 0\,,
\end{equation}
using the fact that for first factor is non-vanishing. Similarly, one has
\begin{subequations}
\begin{align}
[(\vec{j}_2 \times \vec{k}_4 - \vec{j}_4 \times \vec{k}_2) \times \vec{j}_2] \cdot \vec{k}_2 &= \vec{j}_2 \cdot \vec{j}_4\,,\\
[(\vec{j}_3 \times \vec{k}_4 - \vec{j}_4 \times \vec{k}_3) \times \vec{j}_3] \cdot \vec{k}_3 &= \vec{j}_3 \cdot \vec{j}_4\,,\\
[(\vec{j}_2 \times \vec{j}_4 - \vec{k}_2 \times \vec{k}_4) \times \vec{j}_2] \cdot \vec{k}_2 &= \vec{j}_4 \cdot \vec{k}_2\,,\\
[(\vec{j}_3 \times \vec{j}_4 - \vec{k}_3 \times \vec{k}_4) \times \vec{j}_3] \cdot \vec{k}_3 &= \vec{j}_4 \cdot \vec{k}_3\,,
\end{align}
\end{subequations}
and so we arrive at
\begin{equation}
\vec{j}_3 \cdot \vec{k}_4 = \vec{j}_4 \cdot \vec{k}_3 = \vec{j}_2 \cdot \vec{k}_4 = \vec{j}_4 \cdot \vec{k}_2 = \vec{j}_3 \cdot \vec{j}_4 = \vec{k}_3 \cdot \vec{k}_4 = \vec{j}_2 \cdot \vec{j}_4 = \vec{k}_2 \cdot \vec{k}_4 = 0\,.
\end{equation}
Finally, we use these relations to calculate
\begin{subequations}
\begin{align}
0 = (\vec{j}_2 \times \vec{k}_4 - \vec{j}_4 \times \vec{k}_2) \times \vec{j}_2 &= (\vec{j}_2 \cdot \vec{j}_2)\vec{k}_4\,,\\
0 = (\vec{j}_2 \times \vec{j}_4 - \vec{k}_2 \times \vec{k}_4) \times \vec{j}_2 &= (\vec{j}_2 \cdot \vec{j}_2)\vec{j}_4\,,
\end{align}
\end{subequations}
and so we arrive at the ultimate conclusion
\begin{equation}
\vec{j}_4 = \vec{k}_4 = 0\,.
\end{equation}
To summarize, we find the following two homomorphisms:
\begin{enumerate}
\item
If \(\boldsymbol{\lambda}_{1,2,3} = 0\) are given by the trivial homomorphism, \(\boldsymbol{\lambda}_4\) is left arbitrary. Hence, we have \(\vec{j}_1 = \vec{j}_2 = \vec{j}_3 = \vec{k}_1 = \vec{k}_2 = \vec{k}_3 = 0\), and the inequivalent choices of \(\vec{j}_4\) and \(\vec{k}_4\) are described by the orbits of the Lorentz group action on \(\mathfrak{so}(1,3)\) as described in section~\ref{sec:lorentz}.

\item
For the non-trivial case, we have the conditions~\eqref{eq:j2k3j3k3IX}, \eqref{eq:j2j3k2k3IX} and~\eqref{eq:nt23IX}. The latter shows in particular that \(\vec{j}_3 \neq 0\), and we can choose \(\vec{j}_3 \parallel \vec{e}_3\). The same condition shows \(\vec{j}_3 \perp \vec{k}_3\), and so we can choose \(\vec{k}_3 \parallel \vec{e}_2\). The relation between their lengths then implies that they can be written as
\begin{equation}
\vec{j}_3 = \vec{e}_3\cosh u\,, \quad
\vec{k}_3 = \vec{e}_2\sinh u
\end{equation}
for some \(u \in \mathbb{R}\). By performing a boost with \(K_1\), we can always set \(u = 0\), and so \(\vec{j}_3 = \vec{e}_3\) and \(\vec{k}_3 = 0\). The conditions~\eqref{eq:j2k3j3k3IX} and~\eqref{eq:j2j3k2k3IX} then imply \(\vec{j}_2 \perp \vec{e}_3\) and \(\vec{k}_2 \perp \vec{e}_3\), in addition to \(\vec{j}_2 \perp \vec{k}_2\) from~\eqref{eq:nt23IX}. We can thus the remaining freedom to rotate with \(J_3\), which leaves \(\vec{j}_3\) unchanged, to set \(\vec{j}_2 \parallel \vec{e}_2\), and we are forced to have \(\vec{k}_2 \parallel \vec{e}_1\). Now by the same argument as before, we can perform a boost with \(K_3\), which leaves \(\vec{j}_3\) invariant, to set \(\vec{j}_2 = \vec{e}_2\) and \(\vec{k}_2 = 0\). The conditions~\eqref{eq:vekIX123} then imply that \(\vec{j}_1 = \vec{e}_2 \times \vec{e}_3 = \vec{e}_1\) and \(\vec{k}_1 = 0\). Finally, \(\vec{j}_4 = \vec{k}_4 = 0\). We have thus shown that all algebra homomorphisms in this class are equivalent, and provided a simple representative.
\end{enumerate}
The corresponding teleparallel geometries will be derived in section~\ref{ssec:cosmoIX}.

\section{Teleparallel cosmologies}\label{sec:cosmo}
We can now use the Lie algebra homomorphisms constructed in the previous sections in order to derive the most general teleparallel spacetime geometries which are invariant under the corresponding group actions on the equal-time hypersurfaces. Again we discuss each Bianchi type separately. We study Bianchi type II in section~\ref{ssec:cosmoII}, Bianchi type III in section~\ref{ssec:cosmoIII} and Bianchi type IX in section~\ref{ssec:cosmoIX}.

\subsection{Bianchi type II}\label{ssec:cosmoII}
We start with the Bianchi type II, for which we discussed the Lie algebra homomorphisms in section~\ref{ssec:algebraII}. For the symmetry generating vector fields, which satisfy the commutation relations~\eqref{eq:algII123} and~\eqref{eq:algII4}, we can choose the explicit coordinate form
\begin{equation}\label{eq:vecfII}
X_1 = \partial_y\,, \quad
X_2 = \partial_z\,, \quad
X_3 = z\partial_y - \partial_x\,, \quad
X_4 = z\partial_x + \frac{x^2 - z^2}{2}\partial_y - x\partial_z\,.
\end{equation}
In section~\ref{ssec:algebraII} we have found two inequivalent Lie algebra homomorphisms, for which we will now derive the corresponding symmetric tetrads.

\subsubsection{Trivial homomorphism}
The first homomorphism we study consists of the trivial homomorphism for the subalgebra generated by the first three vector fields, supplemented with an arbitrary image for the fourth vector field; explicitly, it is given by the matrices
\begin{equation}\label{eq:homotrivII}
\boldsymbol{\lambda}_1 = \boldsymbol{\lambda}_2 = \boldsymbol{\lambda}_3 = \begin{pmatrix}
0 & 0 & 0 & 0\\
0 & 0 & 0 & 0\\
0 & 0 & 0 & 0\\
0 & 0 & 0 & 0
\end{pmatrix}, \quad
\boldsymbol{\lambda}_4 = \begin{pmatrix}
0 & k_1 & k_2 & k_3\\
k_1 & 0 & -j_3 & j_2\\
k_2 & j_3 & 0 & -j_1\\
k_3 & -j_2 & j_1 & 0
\end{pmatrix}.
\end{equation}
We now need to solve the symmetry condition~\eqref{eq:infisymcondwb} for the most general tetrad \(\theta^a{}_{\mu}\). Inserting the first two vector fields \(X_1\) and \(X_2\) and their corresponding Lorentz algebra elements, we find the condition
\begin{equation}
\partial_y\theta^a{}_{\mu} = \partial_z\theta^a{}_{\mu} = 0\,,
\end{equation}
showing that the tetrad components can depend only on the remaining two coordinates \(t\) and \(x\). Continuing with the vector field \(X_3\), we find the conditions
\begin{equation}
\partial_x\theta^a{}_{t,x,y} = 0\,, \quad
\partial_x\theta^a{}_z - \theta^a{}_y = 0\,.
\end{equation}
It follows that their solution is given by
\begin{equation}
\theta^a{}_{t,x,y} = \tilde{\theta}^a{}_{t,x,y}\,, \quad
\theta^a{}_z = \tilde{\theta}^a{}_z + x\tilde{\theta}^a{}_y\,, \quad
\end{equation}
where \(\tilde{\theta}^a{}_{\mu}\) are functions of time \(t\) only. Finally, we come to the vector field \(X_4\). In this case we find that the symmetry condition~\eqref{eq:infisymcondwb} takes the form of a system of linear algebraic equations, which we can write in the form
\begin{equation}\label{eq:trivcondII}
M^a{}_{\mu b}{}^{\nu}\tilde{\theta}^b{}_{\nu} = 0\,.
\end{equation}
Combining the pairs of Lorentz and spacetime indices into super-indices \(A, B\), we see that this system admits non-vanishing solutions if and only if the determinant of the matrix \(M^A{}_B\) vanishes. By direct calculation we find
\begin{equation}
\det M = S^4(S^2 + D - 1)^2\,,
\end{equation}
using the orbit invariants defined by~\eqref{eq:orbinv}. We can thus easily classify all Lie algebra homomorphisms for which \(M^A{}_B\) becomes degenerate, given by either \(S = 0\) or \(D = 1 - S^2\), so that a non-vanishing tetrad solution exists:
\begin{enumerate}
\item
For the orbit with \(D = S = 0\) represented by \(\vec{j} = \vec{k} = 0\) the condition~\eqref{eq:trivcondII} becomes
\begin{equation}
\tilde{\theta}^a{}_x = \tilde{\theta}^a{}_z = 0\,.
\end{equation}
The resulting tetrad therefore has \(\theta^a{}_x = 0\) and is thus degenerate.
\item
Another orbit with \(D = S = 0\) is represented by \(\vec{j} = \vec{e}_3\) and \(\vec{k} = \vec{e}_2\). Also in this case one finds \(\theta^a{}_x = 0\) and in addition \(\theta^2{}_{\mu} = 0\), so that the tetrad is again degenerate.
\item
A one-parameter family of orbits with \(D = c^2 > 0\) and \(S = 0\) is given by \(\vec{j} = c\vec{e}_3\) and \(\vec{k} = 0\). Again one has a degenerate tetrad with \(\theta^a{}_x = 0\) and \(\theta^1{}_{\mu} = \theta^2{}_{\mu} = 0\).
\item
In analogy to the previous case, one has a one-parameter family with \(D = -c^2 < 0\) choosing \(\vec{j} = 0\) and \(\vec{k} = c\vec{e}_3\). Here one finds \(\theta^a{}_x = 0\) and \(\theta^0{}_{\mu} = \theta^3{}_{\mu} = 0\), which is again degenerate.
\item
Finally, one has a one-parameter of orbits with \(D = 1 - S^2\) and \(S \neq 0\), which thus covers both cases \(S > 0\) and \(S < 0\) listed in section~\ref{sec:lorentz}, given by \(\vec{j} = \vec{e}_3\) and \(\vec{k} = S\vec{e}_3\). The solutions have \(\theta^a{}_t = \theta^a{}_y = 0\) and \(\theta^0{}_{\mu} = \theta^3{}_{\mu} = 0\), and are thus also degenerate.
\end{enumerate}
In summary, we find that for all Lie algebra homomorphisms in this class of solutions the tetrad obtained by solving the symmetry conditions~\eqref{eq:infisymcondwb} is degenerate. Hence, none of them yields a viable teleparallel geometry.

\subsubsection{Non-trivial homomorphism}
We now come to the second branch of solutions, for which we found that there is a unique (up to the adjoint action of the Lorentz group) homomorphism given by the matrix representation
\begin{equation}
\boldsymbol{\lambda}_1 = \begin{pmatrix}
0 & 0 & 0 & 0\\
0 & 0 & 0 & 0\\
0 & 0 & 0 & 0\\
0 & 0 & 0 & 0
\end{pmatrix}, \quad
\boldsymbol{\lambda}_2 = \begin{pmatrix}
0 & 0 & -1 & 0\\
0 & 0 & 0 & 0\\
-1 & 0 & 0 & -1\\
0 & 0 & 1 & 0
\end{pmatrix}, \quad
\boldsymbol{\lambda}_3 = \begin{pmatrix}
0 & 1 & 0 & 0\\
1 & 0 & 0 & 1\\
0 & 0 & 0 & 0\\
0 & -1 & 0 & 0
\end{pmatrix}, \quad
\boldsymbol{\lambda}_4 = \begin{pmatrix}
0 & 0 & 0 & 0\\
0 & 0 & -1 & 0\\
0 & 1 & 0 & 0\\
0 & 0 & 0 & 0
\end{pmatrix}.
\end{equation}
We can then proceed in analogy to the previously studied case, and apply the symmetry conditions for each vector field in turn. Also here it is convenient to start with the vector field \(X_1\), from which we find the condition
\begin{equation}
\partial_y\theta^a{}_{\mu} = 0\,,
\end{equation}
so that the components \(\theta^a{}_{\mu}\) must be independent of \(y\). Similarly, applying the symmetry conditions for \(X_2\) and \(X_3\) fixes the dependence of each component on \(z\) and \(x\), respectively, and the resulting tetrad is parametrized by 16 functions of time \(t\); we do not write these steps explicitly, as they happen to be rather lengthy. Finally, the symmetry condition for \(X_4\) yields a system of linear, algebraic equations, which determine 10 of these parameter functions in terms of the remaining 6. The resulting tetrad can then be parametrized by
\begin{subequations}
\begin{align}
\theta^0 &= [\mathcal{C}_2(1 + x^2 + z^2) + \mathcal{C}_1]\dd t + (\mathcal{C}_3x + \mathcal{C}_4z)\dd x + [\mathcal{C}_6(1 + x^2 + z^2) + \mathcal{C}_5]\dd y\nonumber\\
&\phantom{=}+ \{[\mathcal{C}_6(1 + x^2 + z^2) + \mathcal{C}_5 - \mathcal{C}_4]x + \mathcal{C}_3z\}\dd z\,,\\
\theta^1 &= 2\mathcal{C}_2x\dd t + \mathcal{C}_3\dd x + 2\mathcal{C}_6x\dd y + (2\mathcal{C}_6x^2 - \mathcal{C}_4)\dd z\,,\\
\theta^2 &= 2\mathcal{C}_2z\dd t + \mathcal{C}_4\dd x + 2\mathcal{C}_6z\dd y + (2\mathcal{C}_6xz + \mathcal{C}_3)\dd z\,,\\
\theta^3 &= [\mathcal{C}_2(1 - x^2 - z^2) - \mathcal{C}_1]\dd t - (\mathcal{C}_3x + \mathcal{C}_4z)\dd x + [\mathcal{C}_6(1 - x^2 - z^2) - \mathcal{C}_5]\dd y\nonumber\\
&\phantom{=}+ \{[\mathcal{C}_6(1 - x^2 - z^2) - \mathcal{C}_5 + \mathcal{C}_4]x - \mathcal{C}_3z\}\dd z\,,
\end{align}
\end{subequations}
where \(\mathcal{C}_1, \ldots, \mathcal{C}_6\) are free functions of time. It is helpful to note that its determinant is given by
\begin{equation}
\det\theta = 2(\mathcal{C}_2\mathcal{C}_5 - \mathcal{C}_1\mathcal{C}_6)(\mathcal{C}_3^2 + \mathcal{C}_4^2)\,.
\end{equation}
Further, one finds that the non-vanishing components of the metric~\eqref{eq:metric} are given by
\begin{gather}
g_{tt} = -4\mathcal{C}_1\mathcal{C}_2\,, \quad
g_{ty} = -2(\mathcal{C}_2\mathcal{C}_5 + \mathcal{C}_1\mathcal{C}_6)\,, \quad
g_{tz} = xg_{ty}\,, \quad
g_{xx} = \mathcal{C}_3^2 + \mathcal{C}_4^2\,,\nonumber\\
g_{yy} = -4\mathcal{C}_5\mathcal{C}_6\,, \quad
g_{yz} = xg_{yy}\,, \quad
g_{zz} = \mathcal{C}_3^2 + \mathcal{C}_4^2 - 4\mathcal{C}_5\mathcal{C}_6x^2\,.
\end{gather}
One can still perform coordinate transformations, such as a reparametrization of the time coordinate to eliminate one of these functions. However, we will not discuss this here, and keep the result in its most general form.

\subsection{Bianchi type III}\label{ssec:cosmoIII}
We now come to the teleparallel spacetimes of Bianchi type III. In this case the symmetry generating vector fields must satisfy the commutation relations~\eqref{eq:algIII123} and~\eqref{eq:algIII4}. An explicit form is given by
\begin{equation}
X_1 = \partial_y\,, \quad
X_2 = \partial_z\,, \quad
X_3 = \partial_x - y\partial_y\,, \quad
X_4 = y\partial_x + \frac{1}{2}\left(\frac{e^{-2x}}{1 - n^2} - y^2\right)\partial_y - \frac{ne^{-x}}{1 - n^2}\partial_z\,,
\end{equation}
where \(n \in (0, 1)\) parametrizes a class of different geometries. Together with the algebra homomorphisms found in section~\ref{ssec:algebraIII}, we are thus in the position to determine the most general teleparallel geometries which are invariant under the action of these vector fields.

\subsubsection{Trivial homomorphism}
We start with the first class of Lie algebra homomorphisms, where the subalgebra generated by the vector fields \(X_1, X_3, X_4\) is trivially represented, while the image \(\boldsymbol{\lambda}_2\) of \(X_2\) is arbitrary. The general form of this class of homomorphisms thus has the matrix representation
\begin{equation}\label{eq:homotrivIII}
\boldsymbol{\lambda}_1 = \boldsymbol{\lambda}_3 = \boldsymbol{\lambda}_4 = \begin{pmatrix}
0 & 0 & 0 & 0\\
0 & 0 & 0 & 0\\
0 & 0 & 0 & 0\\
0 & 0 & 0 & 0
\end{pmatrix}, \quad
\boldsymbol{\lambda}_2 = \begin{pmatrix}
0 & k_1 & k_2 & k_3\\
k_1 & 0 & -j_3 & j_2\\
k_2 & j_3 & 0 & -j_1\\
k_3 & -j_2 & j_1 & 0
\end{pmatrix}.
\end{equation}
In order to determine the most general tetrad satisfying the symmetry condition~\eqref{eq:infisymcondwb}, we start with the vector field \(X_1\). Here the condition simply reads
\begin{equation}
\partial_y\theta^a{}_{\mu} = 0\,,
\end{equation}
and so the tetrad components must be independent of \(y\). In the next step we apply the vector field \(X_3\). In this case we obtain the conditions
\begin{equation}
\partial_x\theta^a{}_{t,x,z} = 0\,, \quad
\partial_x\theta^a{}_y = \theta^a{}_y\,.
\end{equation}
These equations are easily solved, and the solution takes the form
\begin{equation}
\theta^a{}_{t,x,z} = \tilde{\theta}^a{}_{t,x,z}\,, \quad
\theta^a{}_y = e^x\tilde{\theta}^a{}_y\,,
\end{equation}
where \(\tilde{\theta}^a{}_{\mu}\) are free functions of \(t\) and \(z\) only. Continuing with the vector field \(X_4\), we find the further conditions
\begin{equation}
\partial_z\tilde{\theta}^a{}_{t,z} = 0\,, \quad
n\partial_z\tilde{\theta}^a{}_y = (1 - n^2)\tilde{\theta}^a{}_x\,, \quad
n\partial_z\tilde{\theta}^a{}_x = n\tilde{\theta}^a{}_z - \tilde{\theta}^a{}_y\,.
\end{equation}
The general solution to these equations takes the form
\begin{gather}
\theta^a{}_{t,z} = \hat{\theta}^a{}_{t,z}\,, \quad
\theta^a{}_x = \hat{\theta}^a{}_x\cos\left(\frac{\sqrt{1 - n^2}}{n}z\right) + \frac{n\hat{\theta}^a{}_z - \hat{\theta}^a{}_y}{\sqrt{1 - n^2}}\sin\left(\frac{\sqrt{1 - n^2}}{n}z\right)\,,\nonumber\\
\theta^a{}_y = \left[(\hat{\theta}^a{}_y - n\hat{\theta}^a{}_z)\cos\left(\frac{\sqrt{1 - n^2}}{n}z\right) + \sqrt{1 - n^2}\hat{\theta}^a{}_x\sin\left(\frac{\sqrt{1 - n^2}}{n}z\right) + n\hat{\theta}^a{}_z\right]e^x\,,
\end{gather}
where now \(\hat{\theta}^a{}_{\mu}\) are functions of time \(t\) only. These functions are finally to be constrained by the symmetry condition for the last vector field \(X_2\), together with the non-trivial element \(\boldsymbol{\lambda}_2\). The resulting equations are a system of linear algebraic equations, which takes the symbolic form
\begin{equation}\label{eq:trivcondIII}
M^a{}_{\mu b}{}^{\nu}\hat{\theta}^b{}_{\nu} = 0\,.
\end{equation}
As for the previously studied geometry, we can consider \(M\) as a matrix with $16 \times 16$ components, whose determinant must vanish in order for the system to possess non-vanishing solutions. By direct calculation one finds
\begin{equation}
\det M = \frac{e^{4x}}{n^8}S^4[n^4(S^2 - D - 1) + n^2(D + 2) - 1]^2\,,
\end{equation}
using the notation~\eqref{eq:orbinv}. This vanishes if either \(S = 0\) or
\begin{equation}\label{eq:nonorthIII}
D = \frac{n^4(1 - S^2) - 2n^2 + 1}{n^2(1 - n^2)}\,.
\end{equation}
Comparing this condition with the orbits of the adjoint action of the Lorentz group on its Lie algebra, we find the following solutions:
\begin{enumerate}
\item
An orbit with \(D = S = 0\) is given by \(\vec{j} = \vec{k} = 0\). Solving the symmetry condition yields a tetrad which has \(\theta^a{}_x = 0\), and is hence degenerate.
\item
The second orbit with \(D = S = 0\) can be represented by \(\vec{j} = \vec{e}_3\) and \(\vec{k} = \vec{e}_2\). This tetrad also has \(\theta^a{}_x = 0\), and also \(\theta^2{}_{\mu} = 0\), and so it is also degenerate.
\item
The one-parameter family of orbits with \(D = c^2 > 0\) and \(S = 0\) generated by the elements \(\vec{j} = c\vec{e}_3\) and \(\vec{k} = 0\) has \(\theta^a{}_x = 0\) and \(\theta^1{}_{\mu} = \theta^2{}_{\mu} = 0\), so also this solution is degenerate.
\item
Similarly to the previous case, the one-parameter family with \(D = -c^2 < 0\) and \(S = 0\) which is obtained from \(\vec{j} = 0\) and \(\vec{k} = c\vec{e}_3\) has \(\theta^a{}_x = 0\) and \(\theta^0{}_{\mu} = \theta^3{}_{\mu} = 0\). This is another degenerate case.
\item
Finally, all remaining orbits admitting non-trivial tetrad solutions are given by solving the condition~\eqref{eq:nonorthIII} for arbitrary \(S \neq 0\). A representative for each orbit is given by
\begin{equation}
\vec{j} = \frac{\sqrt{1 - n^2}}{n}\vec{e}_3\,, \quad
\vec{k} = \frac{nS}{\sqrt{1 - n^2}}\vec{e}_3\,.
\end{equation}
The general solution for the tetrad in this case has \(\theta^a{}_t = \theta^a{}_z = 0\) and \(\theta^0{}_{\mu} = \theta^3{}_{\mu} = 0\). Hence, one once again finds a degenerate solution.
\end{enumerate}
The results given above show that the trivial class of homomorphisms does not yield any viable teleparallel geometries, since all cases lead to degenerate tetrad solutions.

\subsubsection{Non-trivial homomorphism}
We then continue with the second branch of solution, which contains a unique algebra homomorphism (up to equivalence under the adjoint action of the Lorentz group). The representative we found in section~\ref{ssec:algebraIII} reads
\begin{equation}
\boldsymbol{\lambda}_1 = \frac{1}{\sqrt{2}}\begin{pmatrix}
0 & 0 & 1 & 0\\
0 & 0 & 0 & 0\\
1 & 0 & 0 & -1\\
0 & 0 & 1 & 0
\end{pmatrix}, \quad
\boldsymbol{\lambda}_2 = \begin{pmatrix}
0 & 0 & 0 & 0\\
0 & 0 & 0 & 0\\
0 & 0 & 0 & 0\\
0 & 0 & 0 & 0
\end{pmatrix}, \quad
\boldsymbol{\lambda}_3 = \begin{pmatrix}
0 & 0 & 0 & 1\\
0 & 0 & 0 & 0\\
0 & 0 & 0 & 0\\
1 & 0 & 0 & 0
\end{pmatrix}, \quad
\boldsymbol{\lambda}_4 = \frac{1}{\sqrt{2}}\begin{pmatrix}
0 & 0 & 1 & 0\\
0 & 0 & 0 & 0\\
1 & 0 & 0 & 1\\
0 & 0 & -1 & 0
\end{pmatrix}.
\end{equation}
We then solve the symmetry conditions for each of these vector fields. We can start with \(X_2\), where the condition becomes
\begin{equation}
\partial_z\theta^a{}_{\mu} = 0\,,
\end{equation}
and so the components of the tetrad do not depend on \(z\). Similarly, applying the conditions for \(X_1\) and \(X_3\) will fix the (non-trivial) dependence on the remaining spatial coordinates \(x\) and \(y\); we will not display these steps explicitly for brevity. Finally, \(X_4\) gives a system of linear algebraic equation, which determines 10 components. The general solution is thus parametrized by 6 free functions \(\mathcal{C}_1, \ldots, \mathcal{C}_6\) of time, and can be written in the form
\begin{subequations}
\begin{align}
\theta^0 &= \frac{\mathcal{C}_1}{2\sqrt{2}}\left[\frac{e^{-x}}{\sqrt{1 - n^2}} + \sqrt{1 - n^2}(y^2 + 2)e^x\right]\dd t\nonumber\\
&\phantom{=}+ \left\{\left[\frac{e^{-x}}{\sqrt{1 - n^2}} - (y^2 + 2)e^x\right]\frac{\mathcal{C}_2}{2\sqrt{2}}- \frac{\mathcal{C}_3y}{\sqrt{2}\sqrt{1 - n^2}}\right\}\dd x\nonumber\\
&\phantom{=}+ \frac{1}{2\sqrt{2}}\left[\sqrt{1 - n^2}(\mathcal{C}_3 + \mathcal{C}_4)(y^2 + 2)e^{2x} - \frac{\mathcal{C}_3 - \mathcal{C}_4}{\sqrt{1 - n^2}} - 2ye^x\mathcal{C}_2\right]\dd y\nonumber\\
&\phantom{=}+ \frac{\mathcal{C}_4}{2\sqrt{2}n}\left[\frac{e^x}{\sqrt{1 - n^2}} + \sqrt{1 - n^2}(y^2 + 2)e^x\right]\dd z\,,\\
\theta^1 &= \mathcal{C}_5\dd t + \mathcal{C}_6e^x\dd y + \frac{\mathcal{C}_6}{n}\dd z\,,\\
\theta^2 &= -\sqrt{1 - n^2}\mathcal{C}_1ye^x\dd t + \left(\mathcal{C}_2ye^x + \frac{\mathcal{C}_3}{\sqrt{1 - n^2}}\right)\dd x\nonumber\\
&\phantom{=}+ \left[\mathcal{C}_2e^x - \sqrt{1 - n^2}(\mathcal{C}_3 + \mathcal{C}_4)ye^{2x}\right]\dd y - \frac{\sqrt{1 - n^2}}{n}\mathcal{C}_4ye^x\dd z\,,\\
\theta^3 &= \frac{\mathcal{C}_1}{2\sqrt{2}}\left[\frac{e^{-x}}{\sqrt{1 - n^2}} + \sqrt{1 - n^2}(y^2 - 2)e^x\right]\dd t\nonumber\\
&\phantom{=}+ \left\{\left[\frac{e^{-x}}{\sqrt{1 - n^2}} - (y^2 - 2)e^x\right]\frac{\mathcal{C}_2}{2\sqrt{2}}- \frac{\mathcal{C}_3y}{\sqrt{2}\sqrt{1 - n^2}}\right\}\dd x\nonumber\\
&\phantom{=}+ \frac{1}{2\sqrt{2}}\left[\sqrt{1 - n^2}(\mathcal{C}_3 + \mathcal{C}_4)(y^2 - 2)e^{2x} - \frac{\mathcal{C}_3 - \mathcal{C}_4}{\sqrt{1 - n^2}} - 2ye^x\mathcal{C}_2\right]\dd y\nonumber\\
&\phantom{=}+ \frac{\mathcal{C}_4}{2\sqrt{2}n}\left[\frac{e^x}{\sqrt{1 - n^2}} + \sqrt{1 - n^2}(y^2 - 2)e^x\right]\dd z\,.
\end{align}
\end{subequations}
Its determinant is given by
\begin{equation}
\det\theta = (\mathcal{C}_2^2 + \mathcal{C}_3^2)(\mathcal{C}_4\mathcal{C}_5 - \mathcal{C}_1\mathcal{C}_6)\frac{e^x}{n\sqrt{1 - n^2}}\,.
\end{equation}
Finally, one finds that the non-vanishing metric components read
\begin{gather}
g_{tt} = \mathcal{C}_5^2 - \mathcal{C}_1^2\,, \quad
g_{tz} = \frac{\mathcal{C}_5\mathcal{C}_6 - \mathcal{C}_1\mathcal{C}_4}{n}\,, \quad
g_{ty} = ng_{tz}e^x\,, \quad
g_{xx} = \frac{\mathcal{C}_2^2 + \mathcal{C}_3^2}{1 - n^2}\,,\nonumber\\
g_{zz} = \frac{\mathcal{C}_6^2 - \mathcal{C}_4^2}{n^2}\,, \quad
g_{yz} = ng_{zz}e^x\,, \quad
g_{yy} = (\mathcal{C}_2^2 + \mathcal{C}_3^2 - \mathcal{C}_4^2 + \mathcal{C}_6^2)e^{2x}\,.
\end{gather}
Also here we can eliminate one function by a reparametrization of the time coordinate. However, we will not discuss particular coordinate choices here.

\subsection{Bianchi type IX}\label{ssec:cosmoIX}
We finally come to the teleparallel cosmologies of Bianchi type IX, for which we derived the algebra homomorphisms in section~\ref{ssec:algebraIX}. An explicit form of the vector fields, which satisfy the commutation relations~\eqref{eq:algIX123} and~\eqref{eq:algIX4}, is given by
\begin{gather}
X_1 = \partial_y\,, \quad
X_2 = n\frac{\sin y}{\sin x}\partial_z + \cos y\partial_x - \frac{\sin y}{\tan x}\partial_y\,,\nonumber\\
X_3 = n\frac{\cos y}{\sin x}\partial_z - \sin y\partial_x - \frac{\cos y}{\tan x}\partial_y\,, \quad
X_4 = \partial_z\,,
\end{gather}
where \(n \neq 0\) is a constant parameter. Also in this case we have found two classes of Lie algebra homomorphisms, which we will study now in detail.

\subsubsection{Trivial homomorphism}
We start with the class of Lie algebra homomorphisms whose kernel contains the translation algebra generated by the vector fields \(X_1, X_2, X_3\), while the image of the last vector field \(X_4\) is arbitrary, so that the corresponding matrix representation can be written in the form
\begin{equation}\label{eq:homotrivIX}
\boldsymbol{\lambda}_1 = \boldsymbol{\lambda}_2 = \boldsymbol{\lambda}_3 = \begin{pmatrix}
0 & 0 & 0 & 0\\
0 & 0 & 0 & 0\\
0 & 0 & 0 & 0\\
0 & 0 & 0 & 0
\end{pmatrix}, \quad
\boldsymbol{\lambda}_4 = \begin{pmatrix}
0 & k_1 & k_2 & k_3\\
k_1 & 0 & -j_3 & j_2\\
k_2 & j_3 & 0 & -j_1\\
k_3 & -j_2 & j_1 & 0
\end{pmatrix}.
\end{equation}
In order to solve the symmetry conditions, we start with the vector field \(X_1\). The corresponding condition reads
\begin{equation}
\partial_y\theta^a{}_{\mu} = 0\,,
\end{equation}
from which we find that the tetrad components must be independent of \(y\). For the next step, one can consider the linear combination
\begin{equation}\label{eq:condmIX}
[(\mathcal{L}_{X_2}\theta)^a{}_{\mu} + \boldsymbol{\lambda}_2^a{}_b\theta^b{}_{\mu}]\cos y - [(\mathcal{L}_{X_3}\theta)^a{}_{\mu} + \boldsymbol{\lambda}_3^a{}_b\theta^b{}_{\mu}]\sin y = 0\,,
\end{equation}
which yields the conditions
\begin{equation}
\partial_x\theta^a{}_{t,x,z} = 0\,, \quad
\partial_x\theta^a{}_y\sin x = \theta^a{}_y\cos x - n\theta^a{}_z\,.
\end{equation}
The general solution thus takes the form
\begin{equation}
\theta^a{}_{t,x,z} = \tilde{\theta}^a{}_{t,x,z}\,, \quad
\theta^a{}_y = n\tilde{\theta}^a{}_z\cos x + \tilde{\theta}^a{}_y\sin x\,,
\end{equation}
where \(\tilde{\theta}^a{}_{\mu}\) are now functions of \(t\) and \(z\). Similarly, the linear combination
\begin{equation}\label{eq:condpIX}
[(\mathcal{L}_{X_2}\theta)^a{}_{\mu} + \boldsymbol{\lambda}_2^a{}_b\theta^b{}_{\mu}]\sin y + [(\mathcal{L}_{X_3}\theta)^a{}_{\mu} + \boldsymbol{\lambda}_3^a{}_b\theta^b{}_{\mu}]\cos y = 0
\end{equation}
now leads to the conditions
\begin{equation}
\partial_z\tilde{\theta}^a{}_{t,z} = 0\,, \quad
n\partial_z\tilde{\theta}^{}_x = -\tilde{\theta}^a{}_y\,, \quad
n\partial_z\tilde{\theta}^{}_y = \tilde{\theta}^a{}_x\,.
\end{equation}
The solution is straightforward and reads
\begin{equation}
\tilde{\theta}^a{}_{t,z} = \hat{\theta}^a{}_{t,z}\,, \quad
\tilde{\theta}^a{}_x = \hat{\theta}^a{}_x\cos\frac{z}{n} - \hat{\theta}^a{}_y\sin\frac{z}{n}\,, \quad
\tilde{\theta}^a{}_y = \hat{\theta}^a{}_y\cos\frac{z}{n} + \hat{\theta}^a{}_x\sin\frac{z}{n}\,,
\end{equation}
in terms of the integration constants \(\hat{\theta}^a{}_{\mu}\) which depend on time \(t\) only. These are finally solved for from the last symmetry condition given by the vector field \(X_4\) and the non-trivial element \(\boldsymbol{\lambda}_4\). As for the geometries we have studied before, this yields a system of linear algebraic equations of the generic form
\begin{equation}\label{eq:trivcondIX}
M^a{}_{\mu b}{}^{\nu}\hat{\theta}^b{}_{\nu} = 0\,.
\end{equation}
Understanding \(M\) as a square matrix, which must be degenerate in order for non-vanishing tetrad solutions to exist, we can calculate its determinant and find
\begin{equation}
\det M = \frac{\sin^4x}{n^8}S^4(n^4S^2 + n^2D - 1)^2\,,
\end{equation}
using the abbreviations~\eqref{eq:orbinv}. We can thus determine all non-vanishing tetrad solutions as follows:
\begin{enumerate}
\item
The first orbit with \(D = S = 0\) consists of the fixed point \(\vec{j} = \vec{k} = 0\). The most general tetrad found in this case has \(\theta^a{}_x = 0\), and is thus degenerate.
\item
Choosing the orbit with \(D = S = 0\) represented by \(\vec{j} = \vec{e}_3\) and \(\vec{k} = \vec{e}_2\), one also finds \(\theta^a{}_x = 0\), and in addition \(\theta^2{}_{\mu} = 0\). This tetrad is also degenerate.
\item
Turning to the one-parameter family of orbits with \(D = c^2 > 0\) and \(S = 0\) generated by the elements \(\vec{j} = c\vec{e}_3\) and \(\vec{k} = 0\), one finds \(\theta^a{}_x = 0\) and \(\theta^1{}_{\mu} = \theta^2{}_{\mu} = 0\), which is once again degenerate.
\item
Also the one-parameter family with \(D = -c^2 < 0\) and \(S = 0\), which is obtained from \(\vec{j} = 0\) and \(\vec{k} = c\vec{e}_3\), leads to \(\theta^a{}_x = 0\), as well as \(\theta^0{}_{\mu} = \theta^3{}_{\mu} = 0\). Again the tetrad is degenerate.
\item
The remaining orbits which yield a vanishing determinant have \(D = n^{-2} - n^2S^2\) and \(S \neq 0\). Each orbit can be represented by an element \(\vec{j} = n^{-1}\vec{e}_3\) and \(\vec{k} = nS\vec{e}_3\). We find a tetrad which has \(\theta^a{}_t = \theta^a{}_z = 0\) and \(\theta^0{}_{\mu} = \theta^3{}_{\mu} = 0\), and is thus degenerate.
\end{enumerate}
Hence, we find that also for this Bianchi type all tetrad solutions obtained from the trivial homomorphism class are degenerate, and thus cannot constitute viable teleparallel geometries.

\subsubsection{Non-trivial homomorphism}
Next, we come to the second solution for the Lie algebra homomorphism. Here we have found a representative of the unique equivalence class which is given by the standard representation of \(\mathfrak{so}(3)\) for the first three vector fields, while the last vector field is trivially represented. Thus, in matrix form the corresponding Lie algebra homomorphism is expressed as
\begin{equation}
\boldsymbol{\lambda}_1 = \begin{pmatrix}
0 & 0 & 0 & 0\\
0 & 0 & 0 & 0\\
0 & 0 & 0 & -1\\
0 & 0 & 1 & 0
\end{pmatrix}, \quad
\boldsymbol{\lambda}_2 = \begin{pmatrix}
0 & 0 & 0 & 0\\
0 & 0 & 0 & 1\\
0 & 0 & 0 & 0\\
0 & -1 & 0 & 0
\end{pmatrix}, \quad
\boldsymbol{\lambda}_3 = \begin{pmatrix}
0 & 0 & 0 & 0\\
0 & 0 & -1 & 0\\
0 & 1 & 0 & 0\\
0 & 0 & 0 & 0
\end{pmatrix}, \quad
\boldsymbol{\lambda}_4 = \begin{pmatrix}
0 & 0 & 0 & 0\\
0 & 0 & 0 & 0\\
0 & 0 & 0 & 0\\
0 & 0 & 0 & 0
\end{pmatrix}.
\end{equation}
The most simple condition is obtained from the vector field \(X_4\) and reads
\begin{equation}
\partial_z\theta^a{}_{\mu} = 0\,,
\end{equation}
showing that the tetrad components \(\theta^a{}_{\mu}\) do not depend on \(z\). We then proceed with the vector field \(X_1\), which determines the dependence on \(y\), which now becomes non-trivial due to the fact that we have \(\boldsymbol{\lambda}_1 \neq 0\). The symmetry condition~\eqref{eq:infisymcondwb} becomes
\begin{equation}
\partial_y\theta^0{}_{\mu} = 0\,, \quad
\partial_y\theta^1{}_{\mu} = 0\,, \quad
\partial_y\theta^2{}_{\mu} - \theta^3{}_{\mu} = 0\,, \quad
\partial_y\theta^3{}_{\mu} + \theta^2{}_{\mu} = 0\,,
\end{equation}
and is easily solved by
\begin{equation}
\theta^0{}_{\mu} = \tilde{\theta}^0{}_{\mu}\,, \quad
\theta^1{}_{\mu} = \tilde{\theta}^1{}_{\mu}\,, \quad
\theta^2{}_{\mu} = \tilde{\theta}^2{}_{\mu}\cos y + \tilde{\theta}^3{}_{\mu}\cos y\,, \quad
\theta^3{}_{\mu} = \tilde{\theta}^3{}_{\mu}\cos y - \tilde{\theta}^2{}_{\mu}\sin y\,,
\end{equation}
where \(\tilde{\theta}^a{}_{\mu}\) are functions of \(t\) and \(x\) only. To further constrain these functions, we are left with the vector fields \(X_2\) and \(X_3\). It turns out to be convenient to once again consider the linear combination~\eqref{eq:condmIX}, which is a system of linear differential equations in \(x\) only, and thus determines the dependence of the tetrad components on \(x\). We omit this intermediate result here for brevity. Finally, the linear combination~\eqref{eq:condpIX} becomes a system of linear algebraic equations, which determines 10 components of the tetrad. The most general solution to these equations takes the form
\begin{subequations}
\begin{align}
\theta^0 &= \mathcal{C}_1\dd t + n\mathcal{C}_2\cos x\dd y + \mathcal{C}_2\dd z\,,\\
\theta^1 &= \mathcal{C}_3\cos x\dd t - \mathcal{C}_4\sin x\dd x + (n\mathcal{C}_5\cos^2x - \mathcal{C}_6\sin^2x)\dd y + \mathcal{C}_5\cos x\dd z\,,\\
\theta^2 &= \mathcal{C}_3\sin x\sin y\dd t + (\mathcal{C}_4\cos x\sin y - \mathcal{C}_6\cos y)\dd x + \mathcal{C}_5\sin x\sin y\dd z\nonumber\\
&\phantom{=}+ [(n\mathcal{C}_5 + \mathcal{C}_6)\cos x\sin y + \mathcal{C}_4\cos y]\sin x\dd y\,,\\
\theta^3 &= \mathcal{C}_3\sin x\cos y\dd t + (\mathcal{C}_4\cos x\cos y + \mathcal{C}_6\sin y)\dd x + \mathcal{C}_5\sin x\cos y\dd z\nonumber\\
&\phantom{=}+ [(n\mathcal{C}_5 + \mathcal{C}_6)\cos x\cos y - \mathcal{C}_4\sin y]\sin x\dd y\,,
\end{align}
\end{subequations}
where the functions \(\mathcal{C}_1, \ldots, \mathcal{C}_6\) depend on time \(t\) only. We note that its determinant is given by
\begin{equation}
\det\theta = (\mathcal{C}_2\mathcal{C}_3 - \mathcal{C}_1\mathcal{C}_5)(\mathcal{C}_4^2 + \mathcal{C}_6^2)\sin x\,,
\end{equation}
and that the non-vanishing metric components read
\begin{gather}
g_{tt} = \mathcal{C}_3^2 - \mathcal{C}_1^2\,, \quad
g_{tz} = \mathcal{C}_3\mathcal{C}_5 - \mathcal{C}_1\mathcal{C}_2\,, \quad
g_{ty} = ng_{tz}\cos x\,, \quad
g_{xx} = \mathcal{C}_4^2 + \mathcal{C}_6^2\,,\nonumber\\
g_{zz} = \mathcal{C}_5^2 - \mathcal{C}_2^2\,, \quad
g_{yz} = ng_{zz}\cos x\,, \quad
g_{yy} = g_{xx}\sin^2x + n^2g_{zz}\cos^2x\,.
\end{gather}
This concludes our construction of the most general teleparallel spacetime geometry which is invariant under the group action we studied here.

\section{Conclusion}\label{sec:conclusion}
We have constructed the most general teleparallel spacetime geometries which admit a four-dimensional group of motions, i.e., diffeomorphisms which leave the geometry invariant, acting on their spatial hypersurfaces of equal time. Based on the Bianchi classification, these groups belong to the Bianchi types II, III and IX. In order to determine these geometries, we have first constructed all possible Lie algebra homomorphisms from the symmetry algebra to the Lorentz algebra, and found that they fall into two classes, depending on whether their image in the Lorentz algebra is of dimension one or less (which we denoted as trivial) or three (and denoted non-trivial). In particular, we have seen that each of the aforementioned symmetry groups admits exactly one non-trivial homomorphism (up to equivalence by conjugation with the Lorentz group). We have then used these homomorphisms to explicitly construct all possible tetrads, corresponding to teleparallel geometries in the Weitzenböck gauge, which are invariant under the corresponding symmetries. We have found that for all trivial homomorphisms the resulting tetrad is degenerate, and thus does not yield a viable teleparallel geometry, while for the non-trivial homomorphisms a family of non-degenerate and thus viable tetrads exists, which is parametrized by six free functions of time.

As a spin-off result, we have developed and presented a method to determine all homomorphisms from a given symmetry algebra to the Lorentz algebra, by representing the latter in terms of pairs of vectors in \(\mathbb{R}^3\), and expressing their algebra structure through the vector product. We have shown explicitly how to use this method for the symmetry algebras at hand, and we believe that this will be of future use to construct teleparallel spacetimes with other symmetries, and classify the full set of geometries for a given symmetry group.

Using the tetrads we have found, it is now possible to study the dynamics of the corresponding teleparallel Bianchi spacetimes for any given teleparallel gravity theory. Of most interest will be solutions which will approximate an isotropic universe, at least in their metric and observational parameters, in order to accommodate the isotropy of cosmological observations. Besides the background dynamics, also the theory of (linear) perturbations around these Bianchi backgrounds will be of interest, and in particular provide a new approach to tackle the strong coupling problem in teleparallel gravity. We aim to develop and study these frameworks in future research.

\section*{Acknowledgments}
The author gratefully acknowledges the full support by the Estonian Research Council through the Personal Research Funding project PRG356, as well as the European Regional Development Fund through the Center of Excellence TK133 ``The Dark Side of the Universe''.


\begin{thebibliography}{0}
\bibitem{Planck:2018vyg}
N.~Aghanim \textit{et al.} [Planck],
``Planck 2018 results. VI. Cosmological parameters,''
Astron. Astrophys. \textbf{641} (2020), A6
[erratum: Astron. Astrophys. \textbf{652} (2021), C4]
[arXiv:1807.06209 [astro-ph.CO]].

\bibitem{DiValentino:2021izs}
E.~Di Valentino, O.~Mena, S.~Pan, L.~Visinelli, W.~Yang, A.~Melchiorri, D.~F.~Mota, A.~G.~Riess and J.~Silk,
``In the realm of the Hubble tension\textemdash{}a review of solutions,''
Class. Quant. Grav. \textbf{38} (2021) no.15, 153001
[arXiv:2103.01183 [astro-ph.CO]].

\bibitem{CANTATA:2021ktz}
E.~N.~Saridakis \textit{et al.} [CANTATA],
``Modified Gravity and Cosmology: An Update by the CANTATA Network,''
Springer, 2021,
ISBN 978-3-030-83715-0, 978-3-030-83717-4, 978-3-030-83715-0
[arXiv:2105.12582 [gr-qc]].

\bibitem{Heisenberg:2018vsk}
L.~Heisenberg,
``A systematic approach to generalisations of General Relativity and their cosmological implications,''
Phys. Rept. \textbf{796} (2019), 1-113
[arXiv:1807.01725 [gr-qc]].

\bibitem{Aldrovandi:2013wha}
R.~Aldrovandi and J.~G.~Pereira,
``Teleparallel Gravity: An Introduction,''
Springer, 2013,
ISBN 978-94-007-5142-2, 978-94-007-5143-9

\bibitem{Golovnev:2018red}
A.~Golovnev,
``Introduction to teleparallel gravities,''
[arXiv:1801.06929 [gr-qc]].

\bibitem{Hohmann:2022mlc}
M.~Hohmann,
``Teleparallel gravity,''
Lect. Notes Phys. \textbf{1017} (2023), 145-198
[arXiv:2207.06438 [gr-qc]].

\bibitem{Hohmann:2020zre}
M.~Hohmann,
``Complete classification of cosmological teleparallel geometries,''
Int. J. Geom. Meth. Mod. Phys. \textbf{18} (2021) no.supp01, 2140005
[arXiv:2008.12186 [gr-qc]].

\bibitem{Coley:2022qug}
A.~A.~Coley, R.~J.~van den Hoogen and D.~D.~McNutt,
``Symmetric teleparallel geometries,''
Class. Quant. Grav. \textbf{39} (2022) no.22, 22LT01
[arXiv:2205.10719 [gr-qc]].

\bibitem{Bahamonde:2021gfp}
S.~Bahamonde, K.~F.~Dialektopoulos, C.~Escamilla-Rivera, G.~Farrugia, V.~Gakis, M.~Hendry, M.~Hohmann, J.~Levi Said, J.~Mifsud and E.~Di Valentino,
``Teleparallel gravity: from theory to cosmology,''
Rept. Prog. Phys. \textbf{86} (2023) no.2, 026901
[arXiv:2106.13793 [gr-qc]].

\bibitem{Cai:2015emx}
Y.~F.~Cai, S.~Capozziello, M.~De Laurentis and E.~N.~Saridakis,
``f(T) teleparallel gravity and cosmology,''
Rept. Prog. Phys. \textbf{79} (2016) no.10, 106901
[arXiv:1511.07586 [gr-qc]].

\bibitem{Kodama:1984ziu}
H.~Kodama and M.~Sasaki,
``Cosmological Perturbation Theory,''
Prog. Theor. Phys. Suppl. \textbf{78} (1984), 1-166.

\bibitem{Mukhanov:1990me}
V.~F.~Mukhanov, H.~A.~Feldman and R.~H.~Brandenberger,
``Theory of cosmological perturbations. Part 1. Classical perturbations. Part 2. Quantum theory of perturbations. Part 3. Extensions,''
Phys. Rept. \textbf{215} (1992), 203-333.

\bibitem{Malik:2008im}
K.~A.~Malik and D.~Wands,
``Cosmological perturbations,''
Phys. Rept. \textbf{475} (2009), 1-51
[arXiv:0809.4944 [astro-ph]].

\bibitem{Golovnev:2018wbh}
A.~Golovnev and T.~Koivisto,
``Cosmological perturbations in modified teleparallel gravity models,''
JCAP \textbf{11} (2018), 012
[arXiv:1808.05565 [gr-qc]].

\bibitem{Golovnev:2020zpv}
A.~Golovnev and M.~J.~Guzm\'an,
``Foundational issues in f(T) gravity theory,''
Int. J. Geom. Meth. Mod. Phys. \textbf{18} (2021) no.supp01, 2140007
[arXiv:2012.14408 [gr-qc]].

\bibitem{BeltranJimenez:2020fvy}
J.~Beltr\'an Jim\'enez, A.~Golovnev, T.~Koivisto and H.~Veerm\"ae,
``Minkowski space in $f(T)$ gravity,''
Phys. Rev. D \textbf{103} (2021) no.2, 024054
[arXiv:2004.07536 [gr-qc]].

\bibitem{Blagojevic:2020dyq}
M.~Blagojevi\'c and J.~M.~Nester,
``Local symmetries and physical degrees of freedom in $f(T)$ gravity: a Dirac Hamiltonian constraint analysis,''
Phys. Rev. D \textbf{102} (2020) no.6, 064025
[arXiv:2006.15303 [gr-qc]].

\bibitem{Guzman:2019oth}
M.~J.~Guzm\'an and R.~Ferraro,
``Degrees of freedom and Hamiltonian formalism for $f(T)$ gravity,''
Int. J. Mod. Phys. A \textbf{35} (2020) no.02n03, 2040022
[arXiv:1910.03100 [gr-qc]].

\bibitem{BeltranJimenez:2019nns}
J.~Beltr\'an Jim\'enez and K.~F.~Dialektopoulos,
``Non-Linear Obstructions for Consistent New General Relativity,''
JCAP \textbf{01} (2020), 018
[arXiv:1907.10038 [gr-qc]].

\bibitem{Bahamonde:2022ohm}
S.~Bahamonde, K.~F.~Dialektopoulos, M.~Hohmann, J.~Levi Said, C.~Pfeifer and E.~N.~Saridakis,
``Perturbations in non-flat cosmology for f(T) gravity,''
Eur. Phys. J. C \textbf{83} (2023) no.3, 193
[arXiv:2203.00619 [gr-qc]].

\bibitem{Golovnev:2020nln}
A.~Golovnev and M.~J.~Guzman,
``Nontrivial Minkowski backgrounds in $f(T)$ gravity,''
Phys. Rev. D \textbf{103} (2021) no.4, 044009
[arXiv:2012.00696 [gr-qc]].

\bibitem{Li:2023fto}
M.~Li and H.~Rao,
``Irregular universe in the Nieh-Yan modified teleparallel gravity,''
Phys. Lett. B \textbf{841} (2023), 137929
[arXiv:2301.02847 [gr-qc]].

\bibitem{Rodrigues:2012qua}
M.~E.~Rodrigues, M.~J.~S.~Houndjo, D.~Saez-Gomez and F.~Rahaman,
``Anisotropic Universe Models in f(T) Gravity,''
Phys. Rev. D \textbf{86} (2012), 104059
[arXiv:1209.4859 [gr-qc]].

\bibitem{Rodrigues:2014xam}
M.~E.~Rodrigues, A.~V.~Kpadonou, F.~Rahaman, P.~J.~Oliveira and M.~J.~S.~Houndjo,
``Bianchi type-I, type-III and Kantowski-Sachs solutions in $f(T)$ gravity,''
Astrophys. Space Sci. \textbf{357} (2015) no.2, 129
[arXiv:1408.2689 [gr-qc]].

\bibitem{Paliathanasis:2022vux}
A.~Paliathanasis,
``Classical and Quantum Cosmological Solutions in Teleparallel Dark Energy with Anisotropic Background Geometry,''
Symmetry \textbf{14} (2022) no.10, 1974
[arXiv:2209.08817 [gr-qc]].

\bibitem{Amir:2015wja}
M.~J.~Amir and M.~Yussouf,
``Kantowski-Sachs Universe Models in $f(T)$ Theory of Gravity,''
Int. J. Theor. Phys. \textbf{54} (2015) no.8, 2798-2812
[arXiv:1502.00777 [gr-qc]].

\bibitem{Sharif:2011bi}
M.~Sharif and S.~Rani,
``F(T) Models within Bianchi Type I Universe,''
Mod. Phys. Lett. A \textbf{26} (2011), 1657-1671
[arXiv:1105.6228 [gr-qc]].

\bibitem{Fayaz:2014swa}
V.~Fayaz, H.~Hossienkhani, A.~Farmany, M.~Amirabadi and N.~Azimi,
``Cosmology of $f(T)$ gravity in a holographic dark energy and nonisotropic background,''
Astrophys. Space Sci. \textbf{351} (2014), 299-306.

\bibitem{Fayaz:2015yka}
V.~Fayaz, H.~Hossienkhani, A.~Pasqua, M.~Amirabadi and M.~Ganji,
``f(T) theories from holographic dark energy models within Bianchi type I universe,''
Eur. Phys. J. Plus \textbf{130}, no.2, 28 (2015).

\bibitem{Skugoreva:2017vde}
M.~A.~Skugoreva and A.~V.~Toporensky,
``On Kasner solution in Bianchi I $f(T)$ cosmology,''
Eur. Phys. J. C \textbf{78} (2018) no.5, 377
[arXiv:1711.07069 [gr-qc]].

\bibitem{Skugoreva:2019bwt}
M.~A.~Skugoreva and A.~V.~Toporensky,
``Anisotropic cosmological dynamics in $f(T)$ gravity in the presence of a perfect fluid,''
Eur. Phys. J. C \textbf{79} (2019) no.10, 813
[arXiv:1907.12538 [gr-qc]].

\bibitem{Tretyakov:2021cgb}
P.~V.~Tretyakov,
``Bianchi I cosmological solutions in teleparallel gravity,''
Mod. Phys. Lett. A \textbf{37} (2022) no.08, 2250046
[arXiv:2109.14457 [gr-qc]].

\bibitem{Rodrigues:2013iua}
M.~E.~Rodrigues, I.~G.~Salako, M.~J.~S.~Houndjo and J.~Tossa,
``Locally Rotationally Symmetric Bianchi Type-I cosmological model in $f(T)$ gravity: from early to Dark Energy dominated universe,''
Int. J. Mod. Phys. D \textbf{23} (2014), 1450004
[arXiv:1308.2962 [gr-qc]].

\bibitem{Aslam:2013coa}
A.~Aslam, M.~Jamil and R.~Myrzakulov,
``Noether gauge symmetry for the Bianchi type I model in $f(T)$ gravity,''
Phys. Scripta \textbf{88} (2013), 025003
[arXiv:1308.0325 [gr-qc]].

\bibitem{Paliathanasis:2016vsw}
A.~Paliathanasis, J.~D.~Barrow and P.~G.~L.~Leach,
``Cosmological Solutions of $f(T)$ Gravity,''
Phys. Rev. D \textbf{94} (2016) no.2, 023525
[arXiv:1606.00659 [gr-qc]].

\bibitem{Paliathanasis:2017htk}
A.~Paliathanasis, J.~Levi Said and J.~D.~Barrow,
``Stability of the Kasner Universe in f(T) Gravity,''
Phys. Rev. D \textbf{97} (2018) no.4, 044008
[arXiv:1709.03432 [gr-qc]].

\bibitem{Coley:2023ibm}
A.~A.~Coley and R.~J.~van den Hoogen,
``Spatially Homogeneous Teleparallel Gravity: Bianchi I,''
[arXiv:2305.12168 [gr-qc]].

\bibitem{Bianchi:1898}
L.~Bianchi,
``Sugli spazi a tre dimensioni che ammettono un gruppo continuo di movimenti,''
Memorie di Matematica e di Fisica della Societa Italiana delle Scienze, Serie Terza, Tomo XI, pp. 267–352 (1898).

\bibitem{Bianchi:2001}
L.~Bianchi,
``On the Three-Dimensional Spaces Which Admit a Continuous Group of Motions,''
General Relativity and Gravitation \textbf{33} (2001) 2171–2253.

\bibitem{Hohmann:2019nat}
M.~Hohmann, L.~J\"arv, M.~Kr\v{s}\v{s}\'ak and C.~Pfeifer,
``Modified teleparallel theories of gravity in symmetric spacetimes,''
Phys. Rev. D \textbf{100} (2019) no.8, 084002
arXiv:1901.05472 [gr-qc]].

\bibitem{McNutt:2023nxm}
D.~D.~McNutt, A.~A.~Coley and R.~J.~v.~Hoogen,
``A frame based approach to computing symmetries with non-trivial isotropy groups,''
J. Math. Phys. \textbf{64} (2023) no.3, 2881713
[arXiv:2302.11493 [gr-qc]].

\bibitem{Hohmann:2021dhr}
M.~Hohmann,
``A geometric view on local Lorentz transformations in teleparallel gravity,''
Int. J. Geom. Meth. Mod. Phys. \textbf{19} (2022) no.Supp01, 2240001
[arXiv:2112.15173 [gr-qc]].
\end{thebibliography}
\end{document}